\documentclass{IEEEtaes}
\usepackage{color,array,amsthm,amsmath,amssymb,gensymb}
\usepackage{graphicx}
\usepackage{bm}
\usepackage{epstopdf}
\usepackage{subfigure}
\usepackage{epstopdf}
\usepackage{booktabs}
\usepackage{stfloats}
\usepackage{multirow,array}
\usepackage{makecell}

\newcounter{TempEqCnt}
\begin{document}

\title{FDA Jamming Against Airborne Phased-MIMO Radar-Part II: Jamming STAP Performance Analysis}

\author{Yan Sun}

\affil{University of Electronic Science and Technology of China, Chengdu, China} 

\author{Wen-qin Wang}
\member{Senior Member, IEEE}
\affil{University of Electronic Science and Technology of China, Chengdu, China}

\author{Zhou He}

\affil{Southwest Jiaotong University, Chengdu, China} 

\author{Shunsheng Zhang}

\affil{University of Electronic Science and Technology of China, Chengdu, China} 

\receiveddate{~~~~This work was supported by the National Natural Science Foundation of China under Grant 62171092. }

\corresp{{\itshape (Corresponding author: Wen-qin Wang)}.}

\authoraddress{~~~~Y. Sun, W. Wang and S. Zhang are with the School of Information and Communication Engineering, University of Electronic Science and Technology of China, Chengdu 611731, China 
(e-mail: \href{mailto:sunyan_1995@163.com}{sunyan\_1995@163.com}; \href{mailto:wqwang@uestc.edu.cn}{wqwang@uestc.edu.cn}; \href{mailto:zhangss@uestc.edu.cn}{ zhangss@uestc.edu.cn}). Z. He is with the School of Mathematics, Southwest Jiaotong University, Chengdu 611756, China, (e-mail: \href{mailto:zhou.he@swjtu.edu.cn}{zhou.he@swjtu.edu.cn}).}

\maketitle

\begin{abstract}

The first part of this series introduced the effectiveness of frequency diverse array (FDA) jamming through direct wave propagation in countering airborne phased multiple-input multiple-output (Phased-MIMO) radar. This part focuses on the effectiveness of FDA scattered wave (FDA-SW) jamming on the space-time adaptive processing (STAP) for airborne phased-MIMO radar. Distinguished from the clutter signals, the ground equidistant scatterers of FDA-SW jamming constitute an elliptical ring, whose trajectory equations are mathematically derived to further determine the spatial frequency and Doppler frequency. For the phased-MIMO radar with different transmitting partitions, the effects of jamming frequency offset of FDA-SW on the clutter rank and STAP performance are discussed. Theoretical analysis provides the variation interval of clutter rank and the relationship between the jamming frequency offset and the improvement factor (IF) notch of phased-MIMO-STAP. Importantly, the requirements of jamming frequency offset for both two-part applications are discussed in this part. Numerical results verify these mathematical findings and validate the effectiveness of the proposed FDA jamming in countering the phased-MIMO radar.

 
\end{abstract}

\begin{IEEEkeywords}
Phased multiple-input multiple-output (Phased-MIMO) radar, frequency diverse array (FDA), electronic countermeasures (ECM), space-time adaptive processing (STAP), clutter rank.
\end{IEEEkeywords}

\section{Introduction}

The radar detection of moving targets on the airborne platform, particularly in the presence of background clutter and electronic countermeasures (ECM) or jamming [\ref{cite1}], has demanded the development of the space-time adaptive processing (STAP), which has been well-researched in the literature over the past few decades [\ref{cite2}, \ref{cite3}, \ref{cite4}]. With sufficient independent identically distributed (i.i.d.) training data, the conventional STAP techniques can keep the signal-to-clutter-plus-noise radio (SCNR) loss within 3 dB [\ref{cite5}]. From the perspective of ECM, the jamming techniques have also been extensively researched with the booming development of STAP algorithms in the radar community [\ref{cite6}]-[\ref{cite12}], especially two cases of phased multiple-input multiple-output (Phased-MIMO) radar, phased array (PA) radar [\ref{cite6}] and FDA-MIMO radar [\ref{cite7}]-[\ref{cite12}].

Doppler jamming has been widely researched to counter radar tracking and synthetic aperture radar (SAR) imaging. including Doppler deceptive jamming [\ref{cite13}]-[\ref{cite15}], Doppler towing jamming [\ref{cite16}]-[\ref{cite17}], and scattered wave jamming [\ref{cite18}]. The scattered wave jamming was originally proposed for countering synthetic aperture radar (SAR). After intercepting and modulating the radar signal, the scattered wave jamming signal is transmitted to a specific area, then the signal is received by the radar receiver after scattering. With the development of digital radio frequency memory (DRFM) [\ref{cite19}], Doppler-dimensional deception can be effectively achieved by scattered wave jamming. Furthermore, the scattered wave jamming can generate a similar spectrum trajectory with the clutter. Although the target can be separated from the clutter by STAP using the relationship between the spatial frequency and the Doppler frequency, the scattered wave jamming signal can be implemented with prior Doppler information to cover the target in the spatial-Doppler spectrum. Inspired by the published works [\ref{cite18}]-[\ref{cite20}], we believe that the FDA scattered wave (FDA-SW) jamming contributes a more important role against phased-MIMO-STAP by using the additional degrees of freedom (DOFs) provided by the jamming frequency offset.

In our companion paper, the signal model of two types of FDA jamming has been introduced against airborne phased-MIMO radar. The performance of matched-filtering and spatial filtering for the jamming signal directly towards radar has been analyzed. In this second part of the series, we focus on the scenario where the FDA jamming is received by radar through the ground scattered wave propagation. The FDA scattered wave (FDA-SW) jamming and clutter are considered to be suppressed by the two-dimensional (2-D) adaptive filtering technique, STAP [\ref{cite21}]-[\ref{cite23}]. According to the spatial positions of radar and jammer, the trajectory of ground scatterers for FDA-SW jamming signals are mathematically derived, and the space-time signal models of clutter signals and FDA-SW jamming signals are established. Furthermore, the effects of the FDA-SW jamming signals on the clutter rank [\ref{cite24}] and the improvement factor (IF) [\ref{cite25}] notches of phased-MIMO-STAP are analyzed. At last, both analytical and numerical results demonstrate the effectiveness of FDA-SW jamming transmitted by two types of FDA jammer against phased-MIMO radar. The main contribution of this part is briefly summarized as follows. 
 
\begin{enumerate}
\def\labelenumi{\arabic{enumi})}
\item
  We derive the trajectory equation of the ground scatterers for FDA-SW jamming based on the spatial positions of radar and jammer, by which the spatial frequency and Doppler frequency of the FDA-SW jamming signals are determined.
\item
  We mathematically demonstrate the effects of the FDA jamming frequency offset on the clutter rank of phased-MIMO radar with different transmitting partitions. Specifically, the clutter rank of the PA radar is not affected while the clutter rank of the FDA-MIMO radar is affect to increase, and we prove the altered clutter rank intervals.
\item
  We reveal the relationship between the FDA jamming frequency offset and the Doppler-dimensional position of IF notch. Through the theoretical analysis and simulations, adjusting the frequency offset can shift the Doppler-dimensional position of the IF notch, which will effectively protect the target in the Doppler dimension based on the priori information.
\end{enumerate}

The remainder of the paper is organized as follows. The next section contains the derivation of the spatial frequency and Doppler frequency of the FDA-SW jamming signals based on the proposed trajectory equation. Section III introduces the signal models of clutter and FDA-SW jamming. Then the effects of FDA-SW jamming on clutter rank and the IF notch are discussed in Section IV. Section V discusses the requirements of FDA jammer frequency offset for two-part works. Numerical simulation results are presented to demonstrate the effectiveness of FDA-SW jamming on the phased-MIMO-STAP in Section VI. Finally, in Section VII, we draw the conclusions for this series. Derivations are confined to the Appendices. 

The mathematical notation used in this part is fully described in Part I. In this part, we extend the monopulse models in Part I to multipulse models, increasing the Doppler information after matched filtering. We use the superscript '$^{(R)}$' and '$^{(J)}$' to distinguish the clutter signal and FDA-SW jamming signal, respectively.


\begin{figure}[t]
\centerline{\includegraphics[width=19pc]{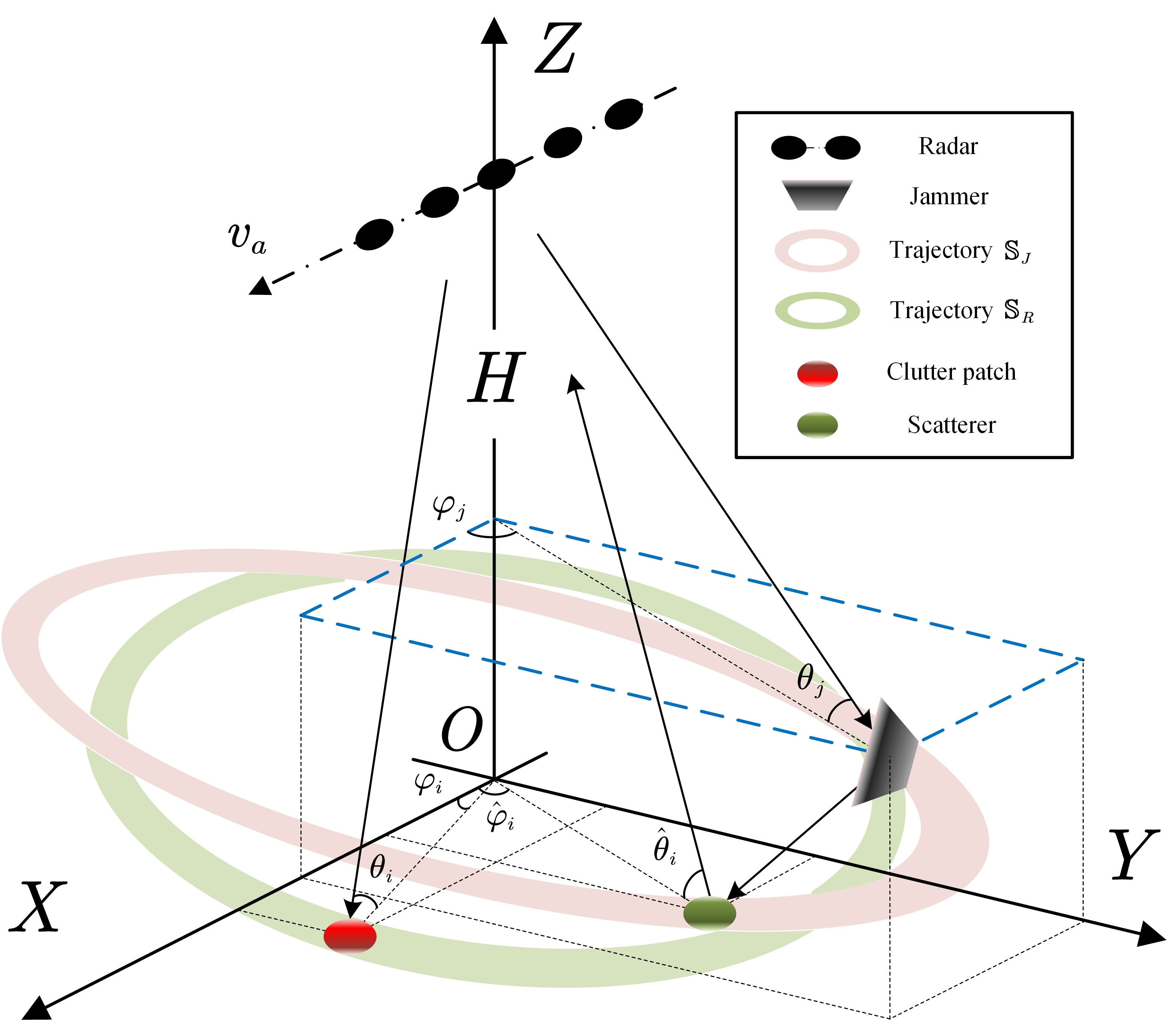}}
\caption{The radar detection scenario of FDA-SW jamming and clutter.}
\label{FIG.1}
\end{figure}

\section{GEOMETRIC SCENARIOS OF DETECTION}

Consider the airborne radar detection scenario in Fig.\ref{FIG.1} of Part I. Assume that the FDA jammer transmits the scattered wave toward the ground, and then airborne phased-MIMO radar receives the jamming signals reflected by the ground, which is similar as the clutter signals. In this part, we focus on echo signals that contain not only target information and noise, but also scattered jamming and clutter signals. To simplify, the expression of `jamming signals' or `jamming' in this part directly represents the scattered wave jamming signals. 

According to the data cube of Pulse Doppler (PD) radar, the target, clutter and scattered wave jamming signals are sampled in the cells under test (CUTs) from the fast-time dimension [\ref{cite1}]. The propagation distance of the jamming (from the jammer to the ground scattering point, and then to the airborne) is assumed to be consistent with the target range (or the distance of clutter signals). Therefore, taking the radar and the jammer as the focus, the propagation distance of jamming as the long axis can be uniquely determined as a spatial ellipsoid, and the intersection line of this ellipsoid with the ground is the distribution trajectory of the ground scattering point for jamming signals, where the points on the trajectory have the same sum of distances to the radar and the jammer. In Fig.\ref{FIG.1}, we have plotted two elliptic ring regions to represent the trajectories of the clutter patches (shown in green) and the ground scatterers for FDA-SW jamming (shown in pink). In the global coordinate system $XYZ$, the coordinates of phased-MIMO radar and FDA jammer can be expressed as $\boldsymbol{p}_{\mathrm{R}}=\left( 0,0,H \right)$ and $\boldsymbol{p}_{\mathrm{J}}=\left( x_{\mathrm{J}},y_{\mathrm{J}},z_{\mathrm{J}} \right)$, then the azimuth $\varphi_j$ and elevation $\theta_j$ of jammer can be expressed as 
\begin{subequations}
  \begin{align}
\varphi _j=&\mathrm{arc}\tan \left( y_{\mathrm{J}}/x_{\mathrm{J}} \right)
\\
  \theta _j=&\mathrm{arc}\tan ( H/\sqrt{x_{\mathrm{J}}^{2}+y_{\mathrm{J}}^{2}})
\end{align}
\label{eq.1a}
\label{eq.1b}
\end{subequations}
And the distance between the radar and the jammer, which is also the focal length of the ellipsoid, can be expressed as
\begin{equation}
R_f=\left\| \boldsymbol{p}_{\mathrm{R}}-\boldsymbol{p}_{\mathrm{J}} \right\|_2
\label{eq.2}
\end{equation}
where $\left\| \cdot \right\| _2$ denotes the L2 norm of vector. Assume that the one-way distance from the target to the radar is $R_t$, then the long axis of the ellipsoid is $2R_t$. The spatial ellipsoid equations and the ground scattering point trajectory equations for the jamming signals are detailed in Appendix A. In the global coordinate system $XYZ$, the trajectory $\mathbb{S} _R$ of the ground scattering point of clutter is a circle centered at the origin, whereas the trajectory $\mathbb{S} _J$ of the ground scattering point of the FDA jamming is determined by (\ref{eq.A7}) in Appendix A.

For the i-th ground scattering point of the FDA-SW jamming with global coordinates of $\hat{\boldsymbol{p}}_i=\left( \hat{x}_i,\hat{y}_i,0 \right)$, the spatial frequency, Doppler frequency and range-dependent frequency can be expressed as 
\begin{subequations}
  \begin{align}
\phi _{s}^{\left( J \right)}=&\frac{df_0}{c}\cos \hat{\varphi}_i\cos \hat{\theta}_i\label{eq.3a}
\\
\phi _{D}^{\left( J \right)}=&\frac{2T}{c}v_a\cos \hat{\varphi}_i\cos \hat{\theta}_i\label{eq.3b}
\\
\phi _{R}^{\left( J \right)}=&\frac{2R_t}{c}\varDelta f\label{eq.3c}
\end{align}
\end{subequations}
where $T$ denotes the pulse repetition interval (PRI). $\hat{\varphi}_i$ and $\hat{\theta}_i$ represents the azimuth and elevation of the i-th scatterer in $\mathbb{S}_J$. Similarly, for the i-th ground scattering point of the clutter with global coordinates of $\boldsymbol{p}_i=\left( x_i,y_i,0 \right)$, the spatial frequency, Doppler frequency and  can be expressed as $\phi _{s}^{\left( R \right)}$, $\phi _{D}^{\left( R \right)}$, and $\phi _{R}^{\left( R \right)}$, which have same formulas as (\ref{eq.3a}), (\ref{eq.3b}), and (\ref{eq.3c}), respectively, but the relevant parameters become $\varphi_i$ and $\theta_i$, where $\varphi_i$ and $\theta_i$ represents the azimuth and elevation of the i-th scatterer in $\mathbb{S}_R$. The range-dependent frequency of the i-th clutter patch can expressed as $\phi _{R}^{\left( R \right)}=\phi _{R}^{\left( J \right)}$.

\begin{subequations}
  \begin{align}
\cos \hat{\varphi}_i=&\frac{\hat{x}_{i}^{2}}{\sqrt{\hat{x}_{i}^{2}+\hat{y}_{i}^{2}}}\label{eq.4a}
\\
\cos \hat{\theta}_i=&\frac{H}{\sqrt{H^2+\left( \hat{x}_{i}^{2}+\hat{y}_{i}^{2} \right)}}\label{eq.4b}
\\
\cos \varphi _i=&\frac{x_{i}^{2}}{\sqrt{x_{i}^{2}+y_{i}^{2}}}\label{eq.4c}
\\
\cos \theta _i=&\frac{\sqrt{R_{t}^{2}-H^2}}{R_t}\label{eq.4d}
\end{align}
\end{subequations}

\section{SIGNAL MODEL OF CLUTTER AND SCATTERED WAVE JAMMING}

This section introduces the clutter signals caused by phased-MIMO radar and the FDA-SW jamming signals caused by two types of FDA jammers, SF jammer and AF jammer. In this part, we expand to multi-pulse signal model.

\subsection{Clutter model}

Focusing on the CUT of the range $R_t$, the clutter signal received by the airborne radar is synthesized from the reflected signals of the clutter patches on the equidistant ring. Assuming that the k-th pulse echo signal of the i-th clutter patches on $\mathbb{S} _R$ is $c_{i,k}^{\left( R \right)}\left( t \right)$ (Referring to (12) in Part I),
\begin{equation}
c_{i,k}^{\left( R \right)}=\xi ^{(R)}_iE(t-\tau^{(R)}_i;\varphi_t ,\theta_t) e^{j2\pi f_0 \left[ (k-1)T+t-\tau ^{(R)}_i\right]}
\label{eq.5}
\end{equation}
where $\xi ^{(R)}_i$ and $\tau ^{(R)}_i$ are the the scattering coefficient and time delay of the i-th clutter patch. Then the synthesized signal of the k-th pulse on the equidistant clutter ring can be expressed as
\begin{equation}
c_{k}^{\left( R \right)}\left( t \right) =\int_0^{2\mathrm{\pi}}{c_{i,k}^{\left( R \right)}\left( t \right)}\mathrm{d}\varphi \approx \sum_{i=1}^{N_c}{c_{i,k}^{\left( R \right)}\left( t \right)}
\label{eq.6}
\end{equation}
where $N_c$ denotes the number of the clutter patches in the equidistant clutter ring. Typically, an equidistant clutter ring can be divided into clutter patches according to equal azimuthal angle, thus converting the integral form into a summation [\ref{cite1}, \ref{cite9}].

\begin{figure}[t]
\centerline{\includegraphics[width=20.2pc]{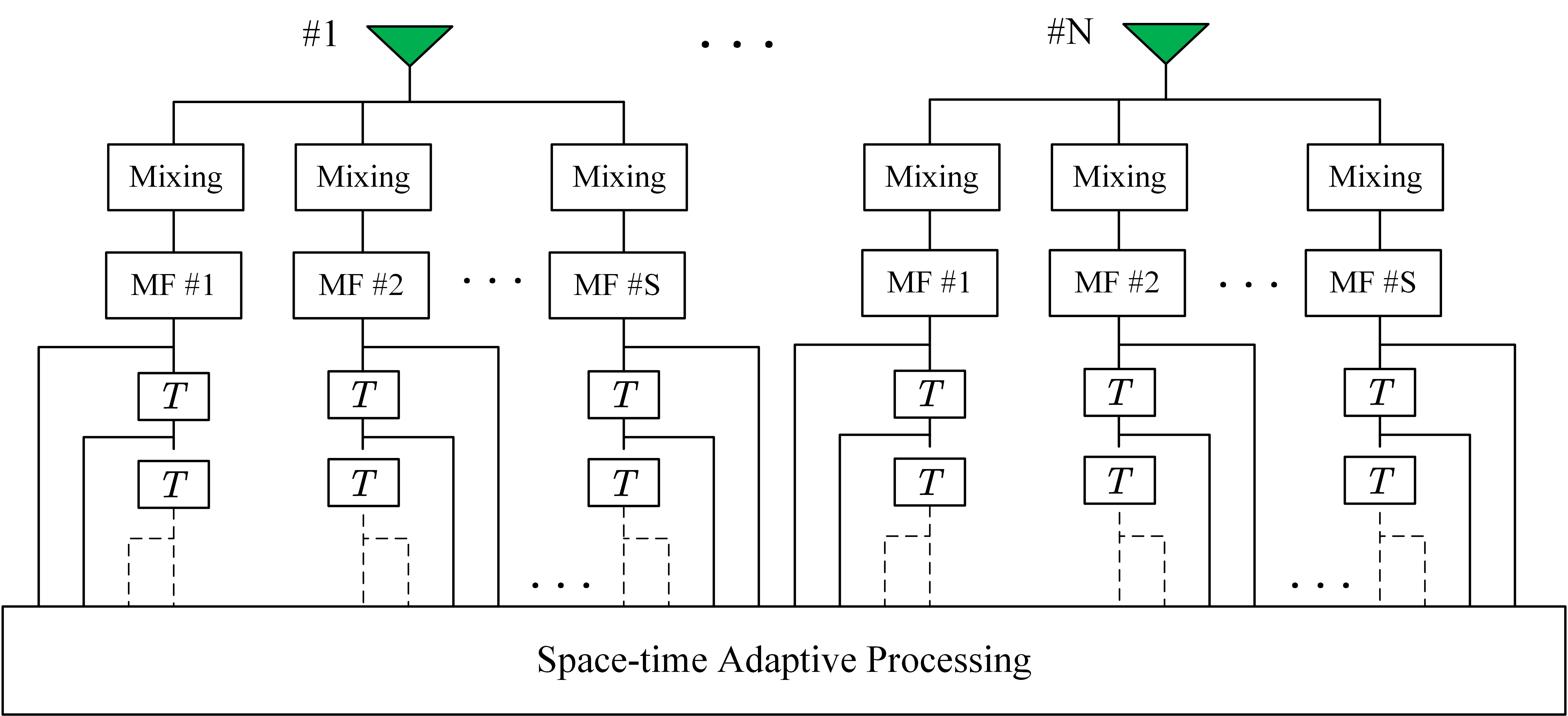}}
\caption{The receiver structure for phased-MIMO-STAP.}
\label{FIG.new}
\end{figure}

For the clutter signal of the i-th clutter patch, after multi-channel mixing and matched-filtering in the radar receiver [\ref{cite1}, \ref{cite3}, \ref{cite8}], whose structure has been shown in Fig.\ref{FIG.new}, the k-th pulse signal of the s-th channel in the n-th receiving array element can be expressed as 
\begin{equation}
\boldsymbol{c}_{i}^{\left( R \right)}=\xi _{i}^{\left( R \right)}\boldsymbol{a}_t\left( \varphi_i ,\theta_i \right) \otimes \boldsymbol{a}_r\left( \varphi_i ,\theta_i \right) \otimes \boldsymbol{d}\left( \varphi_i ,\theta_i  \right) 
\label{eq.7}
\end{equation}
where $\boldsymbol{a}_t\left( \varphi_i ,\theta_i \right)$, $\boldsymbol{a}_r\left( \varphi_i ,\theta_i \right)$, and $\boldsymbol{d}\left( \varphi_i ,\theta_i \right)$ are denote the transmitting subarray spatial frequency vector, the receiving spatial frequency vector and the Doppler frequency vector, respectively [\ref{cite3}, \ref{cite8}]. $\boldsymbol{a}_t\left( \varphi_i ,\theta_i \right) =\boldsymbol{b}\left( \varphi_i ,\theta_i \right) \odot \boldsymbol{c}\left( \varphi_i ,\theta_i \right) \odot \boldsymbol{r}\left( \varDelta f,R_t \right)$ denotes the transmitting steering vector for phased-MIMO radar [\ref{cite3}, \ref{cite34}]. $\boldsymbol{b}\left( \varphi ,\theta \right)$ and $\boldsymbol{c}\left( \varphi ,\theta \right)$ can be refer to (7b) and (7c) in Part I of this series works.
\begin{subequations}
\begin{align}
\boldsymbol{r}\left( \varDelta f,R_t \right) =&\left[ \begin{matrix}
  1&    e^{-j2\mathrm{\pi}\phi _{R}^{\left( R \right)}}&    \cdots&   e^{-j2\mathrm{\pi}\left( S-1 \right) \phi _{R}^{\left( R \right)}}\\
\end{matrix} \right] ^{\mathrm{T}}\label{eq.8a}
\\
\boldsymbol{a}_r\left( \varphi _i,\theta _i \right) =&\left[ \begin{matrix}
  1&    e^{j2\mathrm{\pi}\phi _{s}^{\left( R \right)}}&   \cdots&   e^{j2\mathrm{\pi}\left( N-1 \right) \phi _{s}^{\left( R \right)}}\\
\end{matrix} \right] ^{\mathrm{T}} \label{eq.8b}
\\
\boldsymbol{d}\left( \varphi _i,\theta _i \right) =&\left[ \begin{matrix}
  1&    e^{j2\mathrm{\pi}\phi _{D}^{\left( R \right)}}&   \cdots&   e^{j2\mathrm{\pi}\left( K-1 \right) \phi _{D}^{\left( R \right)}}\\
\end{matrix} \right] ^{\mathrm{T}}\label{eq.8c}
\end{align}
\end{subequations}
where $\phi _{R}^{\left( R \right)}$, $\phi _{s}^{\left( R \right)}$, and $\phi _{D}^{\left( R \right)}$ are referred to (\ref{eq.3a}), (\ref{eq.3b}), and (\ref{eq.3c}), respectively, replacing the relevant parameters with $\phi_i$ and $\theta_i$. For the whole equidistant clutter ring, the clutter signals after matched-filtering can be expressed as $\boldsymbol{c}^{\left( R \right)}=\sum_{i=1}^{N_c}{\boldsymbol{c}_{i}^{\left( R \right)}}$.

\subsection{FDA scattered wave jamming}

In order to simplify the formulas, we express the two FDA jammers uniformly by using superscript $^{(\mathrm{FDA})}$, and explain them separately by using superscript $^{(\mathrm{SF})}$ and $^{(\mathrm{AF})}$ when necessary. FDA-SW deception jamming consists of the reflected echoes from the scattering patches on the equidistant elliptic ring discussed in the previous section. Assuming that the k-th pulse echo signal of the i-th ground scatterer on $\mathbb{S} _J$ is $c_{i,k}^{\left( J \right)}\left( t \right)$ (Referring to (11a) and (11b) in Part I, where $\xi _j$ is modified to the scattering coefficient $\xi ^{J}_i$ of the i-th ground scatterer for FDA-SW jamming), then the synthesized signal of the k-th pulse on the equidistant ring can be expressed as
\begin{equation}
c_{k}^{\left( J \right)}\left( t \right) =\int_{\varphi_{\mathrm{min}}}^{{\varphi_{\mathrm{max}}}}{c_{i,k}^{\left( J \right)}\left( t \right)}\mathrm{d}\varphi \approx \sum_{i=1}^{N_r}{c_{i,k}^{\left( J \right)}\left( t \right)}
\label{eq.9}
\end{equation}
where $N_r$ denotes the number of the scatterers on the equidistant elliptic. $\varphi_{\mathrm{max}}$ and $\varphi_{\mathrm{min}}$ denote the maximum azimuth and minimum azimuth of the scatterers in the trajectory $\mathbb{S}_J$. For the jamming signal of the i-th scatterer patch, after multichannel mixing and matched-filtering, the k-th pulse signal of the s-th channel in the n-th receiving array element can be expressed as
\begin{equation}
\boldsymbol{c}_{i}^{\left( J \right)}=\xi _{i}^{\left( J \right)}\left[ \boldsymbol{\varUpsilon }^{\left( \mathrm{FDA} \right)}\boldsymbol{a}_t\left( \hat{\varphi}_i,\hat{\theta}_i \right) \right] \otimes \boldsymbol{a}_r\left( \hat{\varphi}_i,\hat{\theta}_i \right) \otimes \boldsymbol{d}\left( \hat{\varphi}_i,\hat{\theta}_i \right) 
\label{eq.10}
\end{equation}
where $\xi _{i}^{\left( J \right)}$ represents the scattered coefficient of the i-th scatterer. $\boldsymbol{\varUpsilon }^{\left( \mathrm{FDA} \right)}$ can be consulted in (17b), (17c), (19) and (21) in the Part I. The synthesized  signal from the whole equidistant elliptic can be expressed as $\boldsymbol{c}^{\left( J \right)}=\sum_{i=1}^{N_r}{\boldsymbol{c}_{i}^{\left( J \right)}}$.

\subsection{Clutter-jamming plus noise covariance matrix}
In the fast-time sampled data concerning the CUTs, the signals received by the airborne radar can be expressed as
\begin{equation}
\boldsymbol{y}=\boldsymbol{t}+\boldsymbol{c}^{\left( R \right)}+\boldsymbol{c}^{\left( J \right)}+\boldsymbol{n}
\label{eq.11}
\end{equation}
where $\boldsymbol{t}$ and $\boldsymbol{n}$ denotes the data vectors corresponding to the target and noise, which can be referred to (13) in Part I. Then the clutter-jamming plus noise covariance matrix $\boldsymbol{R}_u$ can be expressed as
\begin{equation}
\boldsymbol{R}_u=\mathrm{E}\left\{ \left( \boldsymbol{c}^{\left( R \right)}+\boldsymbol{c}^{\left( J \right)}+\boldsymbol{n} \right) \left( \boldsymbol{c}^{\left( R \right)}+\boldsymbol{c}^{\left( J \right)}+\boldsymbol{n} \right) ^{\mathrm{H}} \right\} 
\label{eq.12}
\end{equation}
Assuming that the azimuthally divided scatterers on the two trajectories $\mathbb{S}_R$ and $\mathbb{S}_J$ are uncorrelated, then
\begin{equation}
\mathrm{E}\left\{ \left( \xi _{i}^{\left( R \right)} \right) \left( \xi _{j}^{\left( J \right)} \right) ^* \right\} =\,\,\begin{cases}
  \left(\xi _{i}^{\left( R \right)}\right)\left( \xi _{i}^{\left( J \right)} \right) ^*,  \boldsymbol{p}_i=\hat{\boldsymbol{p}}_j\\
   ~~~~~~~~0,     ~~~~~~~~~  \boldsymbol{p}_i\ne \hat{\boldsymbol{p}}_j\\
\end{cases} 
\label{eq.13}
\end{equation}
and the scatterers within each trajectory are also uncorrelated, then
\begin{subequations}
\begin{align}
\mathrm{E}\left\{ \left( \xi _{i}^{\left( c \right)} \right) \left( \xi _{j}^{\left( c \right)} \right) ^* \right\} =&\begin{cases}
  \left| \xi _{i}^{\left( c \right)} \right|^2, i=j\\
  0,      ~~~~~~\;i\ne j\\
\end{cases}\label{eq.14a}
\\
\mathrm{E}\left\{ \left( \xi _{i}^{\left( J \right)} \right) \left( \xi _{j}^{\left( J \right)} \right) ^* \right\} =&\begin{cases}
  \left| \xi _{i}^{\left( J \right)} \right|^2, i=j\\
  0,      ~~~~~~~i\ne j\\
\end{cases}\label{eq.14b}
\end{align}
\end{subequations}
Therefore, the clutter-jamming plus noise covariance matrix $\boldsymbol{R}_u$ can be modified as
\begin{equation}
\boldsymbol{R}_u=\boldsymbol{R}_{c}^{\left( R \right)}+\boldsymbol{R}_{c}^{\left( J \right)}+\boldsymbol{R}_n
\label{eq.15}
\end{equation}
where
\begin{subequations}
\begin{align}
\boldsymbol{R}_{\mathrm{c}}^{\left( R \right)}=&\mathrm{E}\left\{ \left( \boldsymbol{c}^{\left( R \right)} \right) \left( \boldsymbol{c}^{\left( R \right)} \right) ^{\mathrm{H}} \right\} \label{eq.16a}
\\
\boldsymbol{R}_{\mathrm{c}}^{\left( J \right)}=&\mathrm{E}\left\{ \left( \boldsymbol{c}^{\left( J \right)} \right) \left( \boldsymbol{c}^{\left( J \right)} \right) ^{\mathrm{H}} \right\} \label{eq.16b}
\\
\boldsymbol{R}_n=&\mathrm{E}\left\{ \left( \boldsymbol{n} \right) \left( \boldsymbol{n} \right) ^{\mathrm{H}} \right\} \label{eq.16c}
\end{align}
\end{subequations}

\section{EFFECTS OF FDA-SW JAMMING ON PHASED-MIMO-STAP}

This section discusses the effects of scattering wave deceptive interference generated by FDA jammer on phased-MIMO-STAP, in terms of clutter rank and the notches of improvement factor (IF).

\subsection{Clutter rank} 

Let $\xi _{i}^{\left( R \right)}\boldsymbol{v}_{i}^{\left( R \right)}=\boldsymbol{c}_{i}^{\left( R \right)}$ and $\xi _{i}^{\left( J \right)}\boldsymbol{v}_{i}^{\left( J \right)}=\boldsymbol{c}_{i}^{\left( J \right)}$, then the covariance matrices in (\ref{eq.16a}) and (\ref{eq.16b}) can be rewritten as
\begin{subequations}
\begin{align}
\boldsymbol{R}_{\mathrm{c}}^{\left( R \right)}=&\left( \boldsymbol{V}_{\mathrm{c}}^{\left( R \right)} \right) \boldsymbol{\varXi }^{\left( R \right)}\left( \boldsymbol{V}_{\mathrm{c}}^{\left( R \right)} \right) ^{\mathrm{H}}\label{eq.17a}
\\
\boldsymbol{R}_{\mathrm{c}}^{\left( J \right)}=&\left( \boldsymbol{V}_{\mathrm{c}}^{\left( J \right)} \right) \boldsymbol{\varXi }^{\left( J \right)}\left( \boldsymbol{V}_{\mathrm{c}}^{\left( J \right)} \right) ^{\mathrm{H}}\label{eq.17b}
\end{align}
\end{subequations}
where 
\begin{subequations}
\begin{align}
\boldsymbol{V}_{\mathrm{c}}^{\left( R \right)}=&\left[ \begin{matrix}
  \boldsymbol{v}_{1}^{\left( R \right)}&    \boldsymbol{v}_{2}^{\left( R \right)}&    \cdots&   \boldsymbol{v}_{N_c}^{\left( R \right)}\\
\end{matrix} \right]\label{eq.18a}
\\
\boldsymbol{V}_{\mathrm{c}}^{\left( J \right)}=&\left[ \begin{matrix}
  \boldsymbol{v}_{1}^{\left( J \right)}&    \boldsymbol{v}_{2}^{\left( J \right)}&    \cdots&   \boldsymbol{v}_{N_r}^{\left( J \right)}\\
\end{matrix} \right]\label{eq.18b}
\end{align}
\end{subequations}
$\boldsymbol{\varXi }^{\left( R \right)}$ and $\boldsymbol{\varXi }^{\left( J \right)}$ represent the diagonal matrix consisting of scattering coefficients on $\mathbb{S}_R$ and $\mathbb{S}_J$, respectively,
\begin{subequations}
\begin{align}
\boldsymbol{\varXi }^{\left( R \right)}=&\mathrm{diag}\left\{ \left( \xi _{1}^{\left( R \right)} \right) ^2,\left( \xi _{2}^{\left( R \right)} \right) ^2,\cdots ,\left( \xi _{N_c}^{\left( R \right)} \right) ^2 \right\} \label{eq.19a}
\\
\boldsymbol{\varXi }^{\left( J \right)}=&\mathrm{diag}\left\{ \left( \xi _{1}^{\left( J \right)} \right) ^2,\left( \xi _{2}^{\left( J \right)} \right) ^2,\cdots ,\left( \xi _{N_r}^{\left( J \right)} \right) ^2 \right\} \label{eq.19b}
\end{align}
\end{subequations}

The rank of $\boldsymbol{R}_{\mathrm{c}}^{\left( J \right)}$ and $\boldsymbol{R}_{\mathrm{c}}^{\left( J \right)}$ are same as the rank of $\boldsymbol{V}_{\mathrm{c}}^{\left( R \right)}$ and $\boldsymbol{V}_{\mathrm{c}}^{\left( J \right)}$, respectively, since $\boldsymbol{\varXi }^{\left( R \right)}$ and $\boldsymbol{\varXi }^{\left( J \right)}$ are positive definite matrices.
\begin{subequations}
\begin{align}
\mathrm{rank}\left( \boldsymbol{R}_{\mathrm{c}}^{\left( R \right)} \right)  =&\mathrm{rank}\left( \boldsymbol{V}_{\mathrm{c}}^{\left( R \right)} \right) \label{eq.20a} 
\\
\mathrm{rank}\left( \boldsymbol{R}_{\mathrm{c}}^{\left( J \right)} \right)=&\mathrm{rank}\left( \boldsymbol{V}_{\mathrm{c}}^{\left( J \right)} \right)
\label{eq.20b}
\end{align}
\end{subequations}
According to the properties of the matrix rank [\ref{cite24}], it follows that
\begin{align}
\mathrm{rank}\left( \boldsymbol{R}_{\mathrm{c}}^{\left( R \right)}+\boldsymbol{R}_{\mathrm{c}}^{\left( J \right)} \right) &\leqslant \mathrm{rank}\left( \boldsymbol{R}_{\mathrm{c}}^{\left( R \right)} \right) +\mathrm{rank}\left( \boldsymbol{R}_{\mathrm{c}}^{\left( J \right)} \right) 
\nonumber\\
&=\mathrm{rank}\left( \boldsymbol{V}_{\mathrm{c}}^{\left( R \right)} \right) +\mathrm{rank}\left( \boldsymbol{V}_{\mathrm{c}}^{\left( J \right)} \right) 
\label{eq.21}
\end{align}
$\boldsymbol{V}_{\mathrm{c}}^{\left( R \right)}\in \mathbb{C} ^{SNK\times N_c}$ and $\boldsymbol{V}_{\mathrm{c}}^{\left( J \right)}\in \mathbb{C} ^{SNK\times N_r}$ are directly related to the rank of the objective covariance matrix. The i-th column vector $\boldsymbol{v}_{i}^{\left( R \right)}$ and $\boldsymbol{v}_{i}^{\left( J \right)}$ in $\boldsymbol{V}_{\mathrm{c}}^{\left( R \right)}$ and $\boldsymbol{V}_{\mathrm{c}}^{\left( J \right)}$, respectively, are specifically discussed below based on different cases of phased-MIMO radar. The subarray arrangements for the different cases have been described in Part I.

\subsubsection[]{Case 1: PA radar}

Assuming that the transmitting array elements are uniformly weighted, namely, $\left| w_i \right|^2=1, i=1,\cdots ,M$, $\boldsymbol{a}_t\left( \varphi ,\theta \right)$ and $\boldsymbol{\varUpsilon }^{\left( \mathrm{FDA} \right)}\boldsymbol{a}_t\left( \hat{\varphi}_i,\hat{\theta}_i \right)$ in (\ref{eq.7}) and (\ref{eq.10}), respectively, are computed as two scalars $B_{i}^{\left( R \right)}$ and $B_{i}^{\left( J \right)}$.
\begin{subequations}
\begin{align}
B_{i}^{\left( R \right)}=&\frac{\sin \left[ M\mathrm{\pi}\frac{d}{\lambda _0}\left( \cos \varphi _t\cos \theta _t-\cos \varphi _i\cos \theta _t \right) \right]}{\sin \left[ \mathrm{\pi}\frac{d}{\lambda _0}\left( \cos \varphi _t\cos \theta _t-\cos \varphi _i\cos \theta _t \right) \right]}  \label{eq.22a}
\\
B_{i}^{\left( J \right)}=&\varUpsilon ^{\left( \mathrm{FDA} \right)}
\nonumber\\
\times&\frac{\sin \left[ M\mathrm{\pi}\frac{d}{\lambda _0}\left( \cos \varphi _t\cos \theta _t-\cos \hat{\varphi}_i\cos \hat{\theta}_i \right) \right]}{\sin \left[ \mathrm{\pi}\frac{d}{\lambda _0}\left( \cos \varphi _t\cos \theta _t-\cos \hat{\varphi}_i\cos \hat{\theta}_i\right) \right]}\label{eq.22b}
\end{align}
\end{subequations}
where $\varUpsilon ^{\left( \mathrm{FDA} \right)}$ in this case can be referred to (21) in Part I. Then,
\begin{subequations}
\begin{align}
\boldsymbol{v}_{i}^{\left( R \right)}=&B_{i}^{\left( R \right)}\boldsymbol{a}_r\left( \varphi _i,\theta \right) \otimes \boldsymbol{d}\left( \varphi _i,\theta \right) , i=1,\cdots ,N_c \label{eq.23a}
\\
\boldsymbol{v}_{i}^{\left( J \right)}=&B_{i}^{\left( J \right)}\boldsymbol{a}_r\left( \hat{\varphi}_i,\hat{\theta}_i \right) \otimes \boldsymbol{d}\left( \hat{\varphi}_i,\hat{\theta}_i \right) , i=1,\cdots ,N_r
\label{eq.23b}
\end{align}
\end{subequations}

It can be clearly seen that $\boldsymbol{v}_{i}^{\left( J \right)}=\varUpsilon ^{\left( \mathrm{FDA} \right)}\boldsymbol{v}_{i}^{\left( R \right)}$ and $\boldsymbol{V}_{\mathrm{c}}^{\left( J \right)}=\varUpsilon ^{\left( \mathrm{FDA} \right)}\boldsymbol{V}_{\mathrm{c}}^{\left( R \right)}$, so $\boldsymbol{R}_{\mathrm{c}}^{\left( J \right)}=\left| \varUpsilon ^{\left( \mathrm{FDA} \right)} \right|^2\boldsymbol{R}_{\mathrm{c}}^{\left( R \right)}$. Therefore, the FDA jamming cannot affect the clutter rank in this case.
\begin{equation}
\mathrm{rank}\left( \boldsymbol{R}_{\mathrm{c}}^{\left( R \right)}+\boldsymbol{R}_{\mathrm{c}}^{\left( J \right)} \right) =\mathrm{rank}\left( \boldsymbol{R}_{\mathrm{c}}^{\left( R \right)} \right) 
\label{eq.24}
\end{equation}

\subsubsection[]{Case 2: FDA-MIMO radar}
Assuming that the subarrays are uniformly weighted internally and each subarray has the same number of array elements, then $\boldsymbol{b}\left( \varphi ,\theta \right) $ in (7a) of Part I can be rewritten as 
\begin{equation}
\boldsymbol{b}\left( \varphi ,\theta \right) =B_{i}^{\left( R \right)}\left[ \begin{matrix}
  1&    \cdots&   1\\
\end{matrix} \right] ^{\mathrm{T}}
\label{eq.25}
\end{equation}
where
\begin{equation}
B_{i}^{\left( R \right)}=\frac{\sin \left[ M_S\mathrm{\pi}\frac{d}{\lambda _0}\left( \cos \varphi _t\cos \theta _t-\cos \varphi _i\cos \theta _t \right) \right]}{\sin \left[ \mathrm{\pi}\frac{d}{\lambda _0}\left( \cos \varphi _t\cos \theta _t-\cos \varphi _i\cos \theta _t \right) \right]}
\label{eq.26}
\end{equation}
The transmitting steering vector $\boldsymbol{a}_t\left( \varphi ,\theta \right) $ of $\boldsymbol{v}_{i}^{\left( R \right)}$ can be expressed as 
\begin{equation}
\boldsymbol{a}_t\left( \varphi ,\theta \right) =B_{i}^{\left( R \right)}\left[ \boldsymbol{\varUpsilon }^{\left( \mathrm{FDA} \right)}\boldsymbol{p}\left( \varphi ,\theta \right) \right] 
\label{eq.27}
\end{equation}
where 
\begin{align}
&\boldsymbol{p}\left( \varphi ,\theta \right) 
=\left[ \begin{matrix}
  1&        \cdots&   e^{j2\pi \alpha \left( S-1 \right) \left( \alpha \phi _{s}^{\left( R \right)}-\phi _{R}^{\left( R \right)} \right)}\\
\end{matrix} \right] ^{\mathrm{T}}
\label{eq.28}
\end{align}
and $\alpha =\left[ 2\mathrm{\pi}\left( s-1 \right) d\cos \varphi \cos \theta \right] /\left[ \lambda _0\tau _s\left( \varphi ,\theta \right) \right] $. 

In this case, $\boldsymbol{v}_{i}^{\left( R \right)}$ and $\boldsymbol{v}_{i}^{\left( J \right)}$ can be expressed as
\begin{subequations}
\begin{align}
\boldsymbol{v}_{i}^{\left( R \right)}=&B_{i}^{\left( R \right)}\boldsymbol{p}\left( \varphi _i,\theta _t \right) \otimes \boldsymbol{a}_r\left( \varphi _i,\theta _t \right) \otimes \boldsymbol{d}\left( \varphi _i,\theta _t \right) \label{eq.29a}
\\
\boldsymbol{v}_{i}^{\left( J \right)}=&B_{i}^{\left( J \right)}\left[ \boldsymbol{\varUpsilon }^{\left( \mathrm{FDA} \right)}\boldsymbol{p}\left( \hat{\varphi} _i,\hat{\theta} _i \right) \right] 
\otimes \boldsymbol{a}_r\left( \hat{\varphi} _i,\hat{\theta} _i \right)\otimes \boldsymbol{d}\left( \hat{\varphi} _i,\hat{\theta} _i \right) \label{eq.29b}
\end{align}
\end{subequations}
where $B_{i}^{\left( J \right)}$ can be converted to $B_{i}^{\left( R \right)}$ by replacing parameters $\varphi_i$ and $\theta_i$ with $\hat{\varphi} _i$ and $\hat{\theta} _i$, respectively.

Let $z_i=e^{j2\mathrm{\pi}\phi _s\left( \varphi _i,\theta_i \right)}$ and $\beta =2v_aT/d$, the elements in $\boldsymbol{V}_{\mathrm{c}}^{\left( R \right)}$ and $\boldsymbol{V}_{\mathrm{c}}^{\left( J \right)}$ can be rearranged by the fixed indexes $n$, $k$ and $i$.
\begin{subequations}
\begin{align}
\boldsymbol{v}_{i,n,k}^{\left( R \right)}=&z_{i}^{\left( n-1 \right) +\beta \left( k-1 \right)}B_{i}^{\left( R \right)}\boldsymbol{p}\left( \varphi _i,\theta _t \right)  \label{eq.30a}
\\
\boldsymbol{v}_{i,n,k}^{\left( J \right)}=&z_{i}^{\left( n-1 \right) +\beta \left( k-1 \right)}B_{i}^{\left( J \right)}\left[ \boldsymbol{\varUpsilon }^{\left( \mathrm{FDA} \right)}\boldsymbol{p}\left( \hat{\varphi} _i,\hat{\theta} _i \right) \right] \label{eq.30b}
\end{align}
\end{subequations}
Assuming that $Z_{i,n,k}^{\left( R \right)}=B_{i}^{\left( R \right)}z_{i}^{\left( n-1 \right) +\beta \left( k-1 \right)}$ and $Z_{i,n,k}^{\left( J \right)}=B_{i}^{\left( J \right)}z_{i}^{\left( n-1 \right) +\beta \left( k-1 \right)}$, then
\begin{align}
\boldsymbol{V}_{\mathrm{c}}^{\left( J \right)}=&\left[ \begin{matrix}
  \boldsymbol{v}_{1,1,1}^{\left( J \right)}&    \boldsymbol{v}_{2,1,1}^{\left( J \right)}&    \cdots&   \boldsymbol{v}_{N_r,1,1}^{\left( J \right)}\\
  \boldsymbol{v}_{1,2,1}^{\left( J \right)}&    \boldsymbol{v}_{2,2,1}^{\left( J \right)}&    \cdots&   \boldsymbol{v}_{N_r,2,1}^{\left( J \right)}\\
  \vdots&   \vdots&   \ddots&   \vdots\\
  \boldsymbol{v}_{1,N,K}^{\left( J \right)}&    \boldsymbol{v}_{2,N,K}^{\left( J \right)}&    \cdots&   \boldsymbol{v}_{N_r,N,K}^{\left( J \right)}\\
\end{matrix} \right]
\nonumber\\
=&\left( \boldsymbol{I}_{NK}\otimes \boldsymbol{\varUpsilon }^{\left( \mathrm{FDA} \right)} \right) 
\nonumber\\
&\times\left[ \begin{matrix}
  Z_{1,1,1}\boldsymbol{a}_{1}^{\left( t \right)}&   Z_{2,1,1}\boldsymbol{a}_{2}^{\left( t \right)}&   \cdots&   Z_{N_r,1,1}\boldsymbol{a}_{N_r}^{\left( t \right)}\\
  Z_{1,2,1}\boldsymbol{a}_{1}^{\left( t \right)}&   Z_{2,2,1}\boldsymbol{a}_{2}^{\left( t \right)}&   \cdots&   Z_{N_r,2,1}\boldsymbol{a}_{N_r}^{\left( t \right)}\\
  \vdots&   \vdots&   \ddots&   \vdots\\
  Z_{1,N,K}\boldsymbol{a}_{1}^{\left( t \right)}&   Z_{2,N,K}\boldsymbol{a}_{2}^{\left( t \right)}&   \cdots&   Z_{N_r,N,K}\boldsymbol{a}_{N_r}^{\left( t \right)}\\
\end{matrix} \right] 
\nonumber\\
=&\left( \boldsymbol{I}_{NK}\otimes \boldsymbol{\varUpsilon }^{\left( \mathrm{FDA} \right)} \right) \boldsymbol{V}_{\mathrm{c}}^{\left( R \right)}
\label{eq.31}
\end{align}
According to the properties of the matrix rank, we can derive an upper bound of $\mathrm{rank}( \boldsymbol{R}_{\mathrm{c}}^{\left( R \right)}+\boldsymbol{R}_{\mathrm{c}}^{\left( J \right)} )$.
\begin{align}
&\mathrm{rank}\left( \boldsymbol{R}_{\mathrm{c}}^{\left( R \right)}+\boldsymbol{R}_{\mathrm{c}}^{\left( J \right)} \right) 
\nonumber\\
\leqslant &\mathrm{rank}\left( \boldsymbol{R}_{\mathrm{c}}^{\left( R \right)} \right) +\mathrm{rank}\left( \boldsymbol{R}_{\mathrm{c}}^{\left( J \right)} \right) 
\nonumber\\
=&\mathrm{rank}\left( \boldsymbol{V}_{\mathrm{c}}^{\left( R \right)} \right) +\mathrm{rank}\left( \left( \boldsymbol{I}_{NK}\otimes \boldsymbol{\varUpsilon }^{\left( \mathrm{FDA} \right)} \right) \boldsymbol{V}_{\mathrm{c}}^{\left( R \right)} \right) 
\nonumber\\
\leqslant &\min \left\{ \mathrm{rank}\left( \boldsymbol{I}_{NK}\otimes \boldsymbol{\varUpsilon }^{\left( \mathrm{FDA} \right)} \right) ,\mathrm{rank}\left( \boldsymbol{V}_{\mathrm{c}}^{\left( R \right)} \right) \right\} 
\nonumber\\
&+\mathrm{rank}\left( \boldsymbol{V}_{\mathrm{c}}^{\left( R \right)} \right) 
\nonumber\\
\approx &2\left[ S+N-1+\left( K-1 \right) \beta \right] 
\label{eq.32}
\end{align}
And the lower bound of $\mathrm{rank}( \boldsymbol{R}_{\mathrm{c}}^{\left( R \right)}+\boldsymbol{R}_{\mathrm{c}}^{\left( J \right)} )$ can be expressed as
\begin{align}
\mathrm{rank}\left( \boldsymbol{R}_{\mathrm{c}}^{\left( R \right)}+\boldsymbol{R}_{\mathrm{c}}^{\left( J \right)} \right) &\gg \mathrm{rank}\left( \boldsymbol{R}_{\mathrm{c}}^{\left( R \right)} \right) =\mathrm{rank}\left( \boldsymbol{V}_{\mathrm{c}}^{\left( R \right)} \right)
\nonumber\\
&=S+N-1+\left( K-1 \right) \beta
\label{eq.33}
\end{align}

The mathematical discussion demonstrates that FDA jamming can explicitly increase the clutter rank and deteriorate the accuracy of clutter rank estimation for MIMO-STAP. 

\setcounter{TempEqCnt}{\value{equation}} 
\setcounter{equation}{40}
\begin{figure*}[hb]
\hrulefill
\begin{subequations}
\begin{align}
\boldsymbol{\varUpsilon }^{\left( \mathrm{SF} \right)}\boldsymbol{p}\left( \varphi _i,\theta _i \right) &\approx \left[ \begin{matrix}
  e^{-j\left( P-1 \right) \pi \hat{\phi}_{R}^{\left( J \right)}}&   e^{j2\pi \left[ \left( \alpha \phi _{s}^{\left( J \right)}-\phi _{R}^{\left( J \right)} \right) -\frac{\left( P-1 \right)}{2}\hat{\phi}_{R}^{\left( J \right)} \right]}&    \cdots&   e^{j2\pi \left[ \left( S-1 \right) \left( \alpha \phi _{s}^{\left( J \right)}-\phi _{R}^{\left( J \right)} \right) -\frac{\left( P-1 \right)}{2}\hat{\phi}_{R}^{\left( J \right)} \right]}\\
\end{matrix} \right] ^{\mathrm{T}}\label{eq.41a}
\\
\boldsymbol{\varUpsilon }^{\left( \mathrm{AF} \right)}\boldsymbol{p}\left( \hat{\varphi} _i,\hat{\theta} _i \right) &\approx \left[ \begin{matrix}
  e^{j\left( P-1 \right) \pi \left( \phi _{s}^{\left( J \right)}-\hat{\phi}_{R}^{\left( J \right)} \right)}&        \cdots&   e^{j2\pi \left[ \left( S-1 \right) \left( \alpha \phi _{s}^{\left( J \right)}-\phi _{R}^{\left( J \right)} \right) +\frac{\left( P-1 \right)}{2}\left( \phi _{s}^{\left( J \right)}-\hat{\phi}_{R}^{\left( J \right)} \right) \right]}\\
\end{matrix} \right]^{\mathrm{T}}
\label{eq.41b}
\end{align}
\end{subequations}
\end{figure*}
\setcounter{equation}{\value{TempEqCnt}} 
\setcounter{equation}{33}

\subsection{Phased-MIMO-STAP}

The frequency offset attached to the FDA jamming signal can corrupt the phase fitness of the receiving data, which is mentioned in spatial filtering of Part I. For STAP, the jamming frequency offset can also change the linear relationship between the spatial frequency and the Doppler frequency of the side-looking phased-MIMO radar. This section focuses on the effect of the FDA jamming signal on the Doppler dimensional notch of the phased-MIMO-STAP improvement factor (IF).

Assuming that the clutter-jamming plus noise covariance matrix for objective CUT is known, we use the IF to evaluate the STAP performance.
\begin{equation}
\mathrm{IF}=\frac{\left| \boldsymbol{w}_{\mathrm{opt}}^{\mathrm{H}}\boldsymbol{t} \right|^2\sigma _{n}^{2}SNK}{\xi _{t}^{2}\boldsymbol{w}_{\mathrm{opt}}^{\mathrm{H}}\boldsymbol{R}_u\boldsymbol{w}_{\mathrm{opt}}}=\frac{\sigma _{n}^{2}}{\xi _{t}^{2}}\boldsymbol{t}^H\boldsymbol{R}_{u}^{-1}\boldsymbol{t}
\label{eq.34}
\end{equation}
where $\boldsymbol{w}_{\mathrm{opt}}= \boldsymbol{R}_u^{-1}\boldsymbol{t}/\boldsymbol{t}^{\mathrm{H}}\boldsymbol{R}_u^{-1}\boldsymbol{t}$. According to the analysis of Part I, $\boldsymbol{\varUpsilon }^{\left( \mathrm{FDA} \right)}$ has a significant influence on the siganl processing. 

Concerning on $\boldsymbol{\varUpsilon }^{\left( \mathrm{FDA} \right)}=\int_{T_p}{W\left( t;\varDelta f' \right)\boldsymbol{X}\left( t \right) \mathrm{d}t}$, where $\boldsymbol{X}\left( t \right)$ can be refer to (18e) in Part I.
\begin{equation}
W\left( t;\varDelta f' \right) =\left| A\left( t \right) \right|^2\boldsymbol{\rho \vartheta }^{\left( \mathrm{FDA} \right)}\left( t \right) 
\label{eq.35}
\end{equation}
then we assume that $A\left( t \right)$ denotes the rectangular envelopr function with unit energy of time width $T_p$, $\left| A\left( t \right) \right|^2={1}/{T_p}$, and each radial antenna (which has shown in Fig.5 of Part I) of FDA jammer is uniformly weighted, $\boldsymbol{\rho }=\left[ \begin{matrix}
  \xi _{j}^{2}&   \xi _{j}^{2}&   \cdots&   \xi _{j}^{2}\\
\end{matrix} \right] $. We can obtain $W\left( t;\varDelta f' \right)$ for two FDA jammers. 
\begin{subequations}
\begin{align}
W^{\left( \mathrm{SF} \right)}\left( t;\varDelta f' \right) =&\varTheta ^{\left( \mathrm{SF} \right)}\left( \varDelta f' \right) \varPsi ^{\left( \mathrm{SF} \right)}\left( t;\varDelta f' \right) \label{eq.36a}
\\
W^{\left( \mathrm{AF} \right)}\left( t;\varDelta f' \right) =&\varTheta ^{\left( \mathrm{AF} \right)}\left( \varDelta f' \right) \varPsi ^{\left( \mathrm{AF} \right)}\left( t;\varDelta f' \right) \label{eq.36b}
\end{align}
\end{subequations}
where
\begin{subequations}
\begin{align}
\varTheta ^{\left( \mathrm{SF} \right)}\left( \varDelta f' \right) =&\frac{\xi _{j}^{2}}{T_p}e^{-j\left( P-1 \right) \pi \hat{\phi}_{R}^{\left( J \right)}}\label{eq.37a}
\\
\varTheta ^{\left( \mathrm{AF} \right)}\left( \varDelta f' \right) =&\frac{\xi _{j}^{2}}{T_p}e^{j\left( P-1 \right) \pi \left( \phi _{s}^{\left( J \right)}-\hat{\phi}_{R}^{\left( J \right)} \right)}\label{eq.37b}
\\ 
\varPsi ^{\left( \mathrm{SF} \right)}\left( t;\varDelta f' \right) =&e^{j\left( P-1 \right) \pi \varDelta f't}\label{eq.37c}
\nonumber\\
&\times\frac{\sin P\pi \left( \varDelta f't-\hat{\phi}_{R}^{\left( J \right)} \right)}{\sin \pi \left( \varDelta f't-\hat{\phi}_{R}^{\left( J \right)} \right)}
\\
\varPsi ^{\left( \mathrm{AF} \right)}\left( t;\varDelta f' \right) =&e^{j\left( P-1 \right) \pi \varDelta f't}\nonumber\\
&\times\frac{\sin P\pi \left( \varDelta f't-\hat{\phi}_{R}^{\left( J \right)}+\phi _{s}^{\left( J \right)} \right)}{\sin \pi \left( \varDelta f't-\hat{\phi}_{R}^{\left( J \right)}+\phi _{s}^{\left( J \right)} \right)}\label{eq.37d}
\end{align}
\end{subequations}
It should be noted that $\hat{\phi}_{R}^{\left( j \right)}=-\varDelta f'{2R_t}/{c}$ represents the range-dependent frequency caused by $\varDelta f'$.

Using the superscript $^{\left( \mathrm{FDA} \right)}$ to unify the FDA jamming, then $\boldsymbol{\varUpsilon }^{\left( \mathrm{FDA} \right)}$ can be expressed as
\begin{align}
\boldsymbol{\varUpsilon }^{\left( \mathrm{FDA} \right)}=&\varTheta ^{\left( \mathrm{FDA} \right)}\left( \varDelta f' \right)
\nonumber\\
 \times &\left[ \begin{matrix}
   \varOmega _{0}^{\left( \mathrm{FDA} \right)}\left( \varDelta f,\varDelta f'  \right)&    \cdots&   \varOmega _{S-1}^{\left( \mathrm{FDA} \right)}\left( \varDelta f,\varDelta f'  \right)\\
  \vdots&   \ddots&   \vdots\\
  \varOmega _{1-S}^{\left( \mathrm{FDA} \right)}\left( \varDelta f,\varDelta f' \right)&   \cdots&   \varOmega _{0}^{\left( \mathrm{FDA} \right)}\left( \varDelta f,\varDelta f' \right)\\
\end{matrix} \right]
\nonumber\\
=&\varTheta ^{\left( \mathrm{FDA} \right)}\left( \varDelta f' \right) \boldsymbol{D}\left( \varDelta f, \varDelta f' \right) 
\label{eq.38}
\end{align}
where the element of $\boldsymbol{D}\left( \varDelta f, \varDelta f' \right)$ for SF jamming and AF jamming can be expressed as, respectively,
\begin{subequations}
\begin{align}
&\varOmega _{s}^{\left( \mathrm{SF} \right)}\left( s\varDelta f,\varDelta f' \right) 
\nonumber\\
=&\int_{T_p}{\frac{\sin P\pi \left( \varDelta f't-\hat{\phi}_{R}^{\left( J \right)} \right)}{\sin \pi \left( \varDelta f't-\hat{\phi}_{R}^{\left( J \right)} \right)}e^{j2\pi \left( s\varDelta f+\frac{\left( P-1 \right)}{2}\varDelta f' \right) t}\mathrm{d}t}
\label{eq.39a}
\\
&\varOmega _{s}^{\left( \mathrm{AF} \right)}\left( s\varDelta f,\varDelta f' \right) 
\nonumber\\
=&\int_{T_p}{\frac{\sin P\pi \left( \varDelta f't-\hat{\phi}_{R}^{\left( J \right)}+\phi _{s}^{\left( J \right)} \right)}{\sin \pi \left( \varDelta f't-\hat{\phi}_{R}^{\left( J \right)}+\phi _{s}^{\left( J \right)} \right)}e^{j2\pi \left( s\varDelta f+\frac{\left( P-1 \right)}{2}\varDelta f' \right) t}\mathrm{d}t}
\label{eq.39b}
\end{align}
\end{subequations}
where the mathematical discussion about $\boldsymbol{D}\left( \varDelta f, \varDelta f' \right)$ can be seen in Proposition.1. 

\textbf{Proposition 1}\label{Pro.1}: When $\varDelta f\geqslant {1}/{Tp}\gg \varDelta f'$, $\boldsymbol{D}\left( \varDelta f, \varDelta f' \right)$ is the diagonal dominance matrix, which can be approximately represented as

\begin{align}
\boldsymbol{D}\left( \varDelta f, \varDelta f' \right) \approx PT_p\boldsymbol{I}_S
\label{eq.40}
\end{align}

\setcounter{TempEqCnt}{\value{equation}} 
\setcounter{equation}{41}
\textbf{Proof}: See in Appendix B.

Consequently, the transmitting steering vector affected by FDA jamming can be expressed in (\ref{eq.41a}) and (\ref{eq.41b}). In terms of the transmitting steering vector, $\varDelta f'$ induces a shift in the transmitting spatial frequency. For example, the Doppler-dimensional notch of IF that is supposed to be at 90 degrees azimuth (where the Doppler frequency is 0 for the side-looking array) will be shifted, where the amount of Doppler shift caused by two FDA jammers can be expressed as 
\begin{subequations}
\begin{align}
f_{D}^{\left( \mathrm{SF} \right)}=&-\frac{\left( P-1 \right) \hat{\phi}_{R}^{\left( J \right)}}{2\beta}\label{eq.42a}
\\
f_{D}^{\left( \mathrm{AF} \right)}=&\frac{\left( P-1 \right) \left( \phi _{s}^{\left( J \right)}-\hat{\phi}_{R}^{\left( J \right)} \right)}{2\beta}\label{eq.42b}
\end{align}
\end{subequations}


\section{DISCUSSION OF JAMMER FREQUENCY OFFSET}

Noted that the jamming frequency offset is important for the detection scenarios of the proposed FDA jamming in both two parts of this series. This section discusses the working scope of the jamming frequency offset for different functions and scenarios, namely, matched-filtering and spatial filtering in Part I, jamming STAP in Part II. Combined with the conclusions of the jamming frequency offset in Part I, we summarize the all conditions for jamming frequency offset in this section.



\subsection{Range-dimensional dense false targets}

FDA jammers can be used as supportive jamming to generate false targets in the range dimension, as mentioned in Part I. On the one hand, the frequency offset cannot be too large because the frequency offset of the P antennas cannot exceed  the signal bandwidth causing the failure of matched-filtering, namely, $\varDelta f'< \varDelta f/(P-1)$. On the other hand, the frequency offset cannot be too small because the each generated false targets should appear in different range cell, namely, $\varDelta f' \geqslant 1/[(P-1)T_p]$. Therefore, for the supportive jamming after matched-filtering, the FDA jamming frequency offset should be set as
\begin{equation}
\frac{1}{(P-1)T_p}\leqslant\varDelta f'<\frac{\varDelta f}{P-1}
\label{eq.43}
\end{equation}
where $T_p$ is pulse width and $P$ is the number of jammer antennas. Noted that this jamming frequency offset condition is suitable for both SF and AF jammer.

\subsection{Sidelobe \& mainlobe deceptive jamming}

FDA jammers can be used as self-protection jamming to deploy in the sidelobe or mainlobe of radar, which is similar to the traditional sidelobe or mainlobe deceptive jamming. According to the theoretical analysis in Part I, the jamming frequency offsets of SF and AF jammer has the same condition against the spatial filtering of phased-MIMO radar. Therefore, for the self-protection jamming against spatial filtering, the FDA jamming frequency offset should be set as
\begin{equation}
0<\varDelta f'\leqslant \frac{1}{(P-1)T_p}
\label{eq.44}
\end{equation}
Noted that the performance of spatial filtering of phased-MIMO radar deteriorates as the increasing of the jamming frequency offset. 



\subsection{Scattered wave jamming}

FDA jammers can be used as scattered wave jamming against phased-MIMO-STAP, deteriorating the performance of clutter suppression and protecting the target. According to the derivations in this part, the phase terms associated with the frequency offset of the SF jamming and AF jamming are periodic. Therefore, there will be two consequences of the increasing jamming frequency offset. Firstly, the energy of jamming signals will drop severely after MF. Secondly, the features of jamming effect will be repeated. For jamming STAP, the jamming frequency offset of two FDA jammers should preferably satisfy
\begin{subequations}
\begin{align}
\varDelta f'^{\left( \mathrm{SF} \right)}&\leqslant \zeta ^{\left( \mathrm{SF} \right)}=\frac{\beta c}{\left( P-1 \right) R_t}
\label{eq.45a}
\\
\varDelta f'^{\left( \mathrm{SF} \right)}&\leqslant \zeta ^{\left( \mathrm{AF} \right)}=\frac{\beta c}{\left( P-1 \right) R_t}+\frac{\beta f_0d_j}{2R_t}\cos \hat{\theta}_j\cos \hat{\varphi}_j
\label{eq.45b}
\end{align}
\end{subequations}

The requirements of FDA jamming summarized in this section are verified in the simulation results in the next section. More application possibilities for FDA jamming are prospected in section VII.

\begin{table}[t]
 \centering
 \caption{Simulation Parameters}
 \label{tab.2}
 \begin{tabular}{lll}
  \toprule
 Parameter & Symbol & Value  \\
  \midrule
  Reference carrier frequency & ~~$f_0$ & 10~GHz \\
  Platform height & ~~$H$ & 2000~m\\
  Pulse repetition interval & ~~$T$ & 100~us \\
  Platform velocity & ~~$v$ & 75~m/s\\
  Radar antenna spacing & ~~$d$ & 15~mm \\
  Pulse duration & ~~$T_p$ & 10~us\\
  Bandwidth~/~Frequency offset & ~~$B =\Delta f$&   10~$\mathrm{MHz}$\\
  Number of jammer antennas & ~~$P$ & 4 \\
  Number of transmitting antennas & ~~$M$ & 8\\
  Number of receiving antennas & ~~$N$ & 8\\
  Number of pulses & ~~$K$ & 8\\
  Azimuth of target & ~~$\varphi_t $ & 90$^\circ$ \\
  Normalized Doppler frequency of target  & ~~$\phi _{D}^{\left( R \right)}$ &0.25 \\
  Range of target & ~~$R_t$ & 6~km\\
  INR & ~~${\xi _{t}^{2}}/{\sigma _{n}^{2}} $ & 30~$\mathrm{dB}$ \\
  SNR & ~~${\xi _{j}^{2}}/{\sigma _{n}^{2}} $ & 10~$\mathrm{dB}$ \\

  \bottomrule
 \end{tabular}
\end{table}

\begin{figure}[t]
\centerline{\includegraphics[width=20pc]{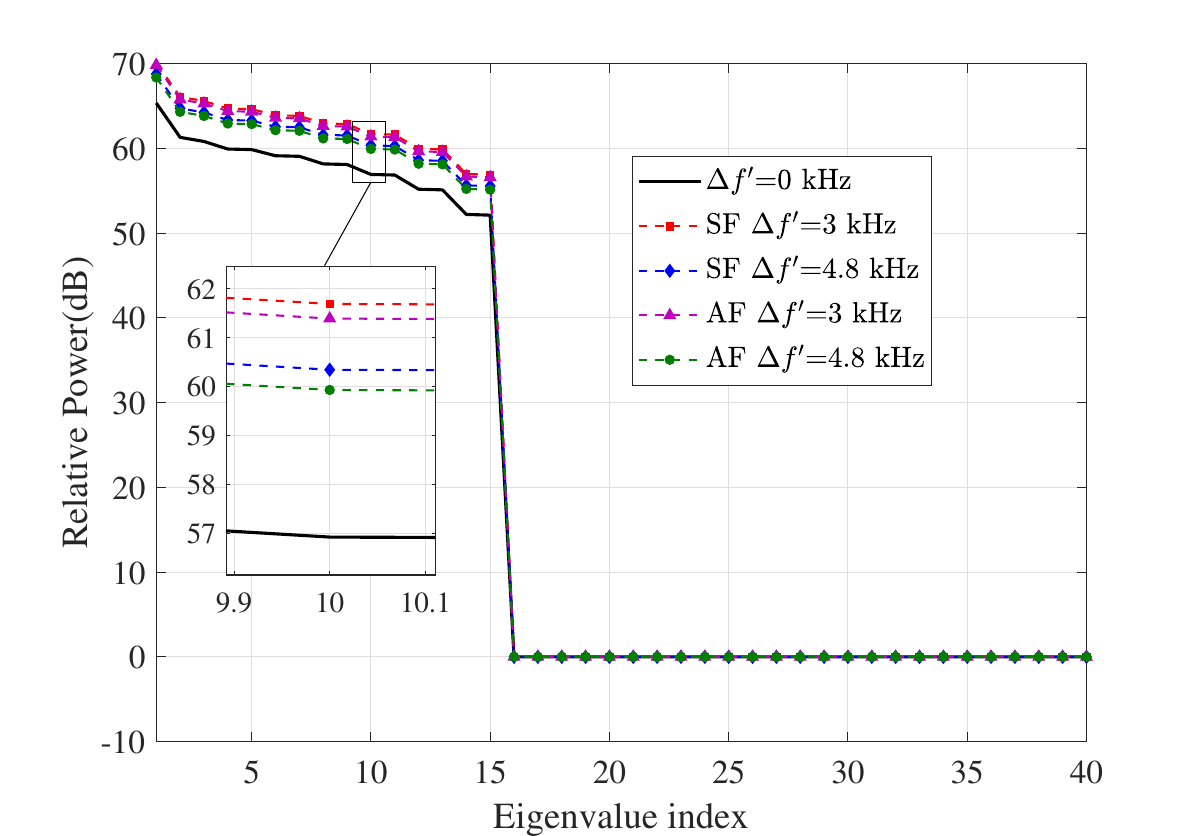}}
\caption{Effect of $\Delta f'$ for FDA jamming on the clutter rank of PA radar.}
\label{FIG.2}
\end{figure}

\begin{figure}[t]
\centerline{\includegraphics[width=20pc]{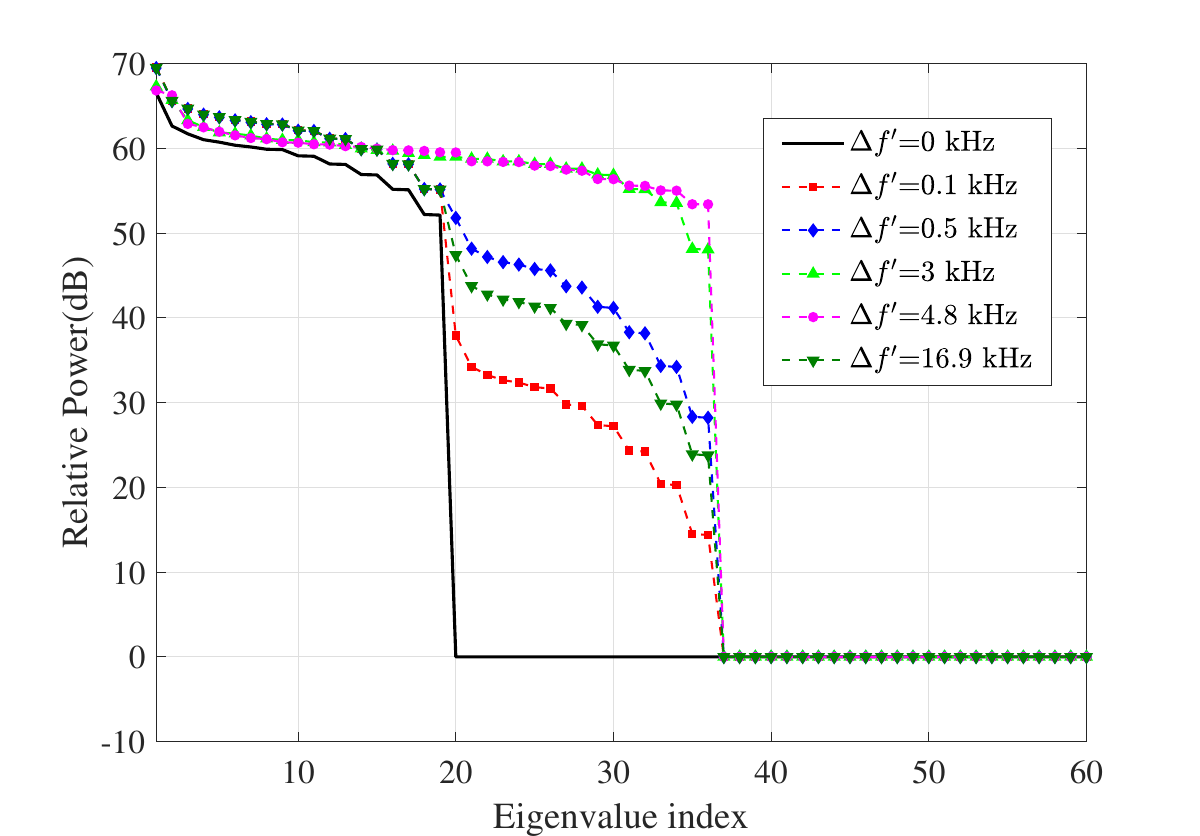}}
\caption{Effect of $\Delta f'$ for SF jamming on the clutter rank of FDA-MIMO radar.}
\label{FIG.3}
\end{figure}

\section{SIMULATION RESULTS}

In this section, numerical simulations are exhibited to verify the effect of FDA jamming on the rank of clutter, spatial-Doppler spectrum and IF for phased-MIMO-STAP. The simulation parameters of phased-MIMO radar, FDA jammer and the target are listed in Table \ref{tab.2}. Noted that according to the calculation of the simulation parameters, $\zeta ^{\left( \mathrm{SF} \right)}\approx 16.7~\mathrm{kHz}$, $\zeta ^{\left( \mathrm{AF} \right)}\approx 18.5~\mathrm{kHz}$, and $\beta = 1$.

We use two kinds of FDA jammers to deteriorate the received signals of PA radar with one subarray ($S=1$) and FDA-MIMO radar with four subarrays ($S=4$), respectively. Fig.\ref{FIG.2}, Fig.\ref{FIG.3}, and Fig.\ref{FIG.4} present the 
clutter eigen-spectrum. Fig.\ref{FIG.5(a)} - Fig.\ref{FIG.5(h)} show the spatial-Doppler spectrum. Fig.\ref{FIG.6} - Fig.\ref{FIG.9} illustrate the simulations of IF. As a comparison, $\varDelta f'=0~\mathrm{kHz}$ indicates that there is no FDA jamming affects the phased-MIMO-STAP.

\subsection{Clutter rank}

Fig.\ref{FIG.2} illustrates the effects of SF jamming and AF jamming with different frequency offsets ($\Delta f'$) on the clutter rank of PA radar. FDA jammers cannot affect the clutter rank of PA radar, which verifies the deduction in (\ref{eq.24}). FDA jammers can increase the clutter eigenvalue of PA radar. In particular, the enhancement of clutter eigenvalue will be smaller when the frequency offset of jamming becomes larger from the zoomed-in perspective. Moreover, the AF jammer requires a larger frequency offset to achieve the same performance as the SF jammer.

\begin{figure}[t]
\centerline{\includegraphics[width=20pc]{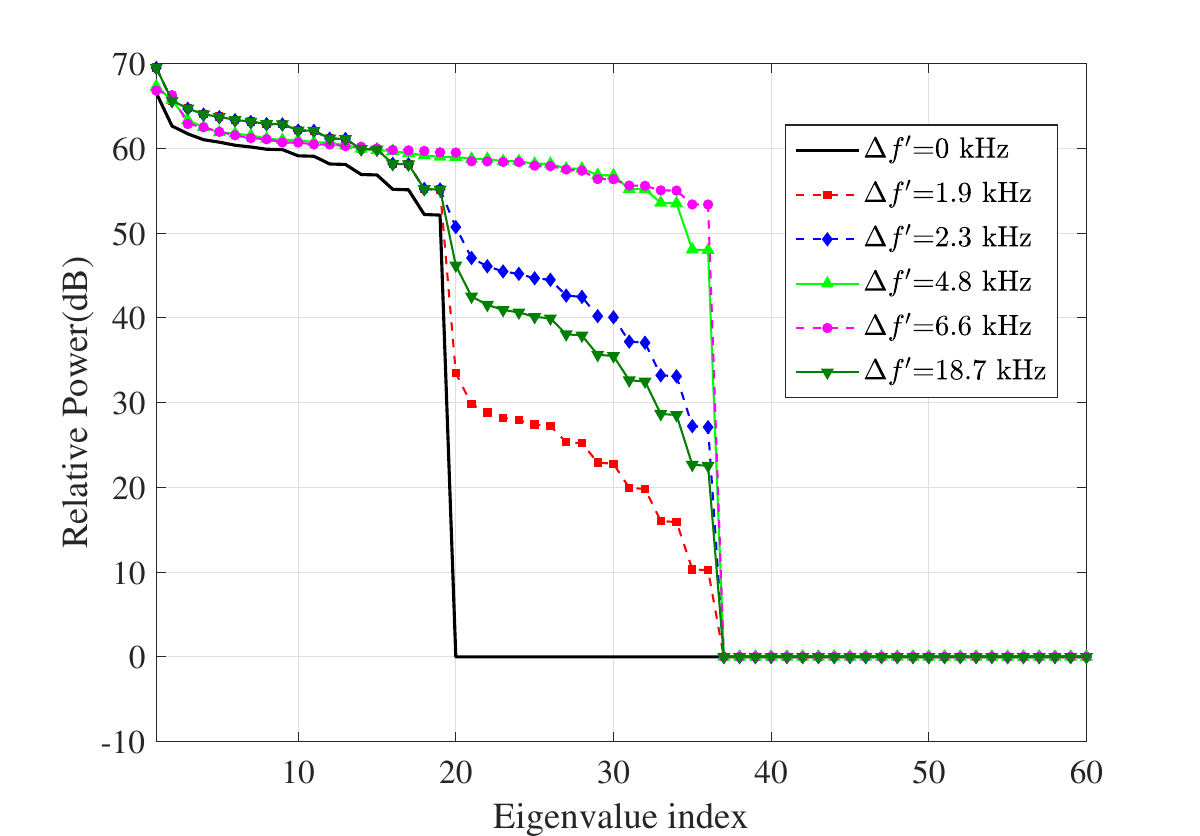}}
\caption{Effect of $\Delta f'$ for AF jamming on the clutter rank of FDA-MIMO radar.}
\label{FIG.4}
\end{figure}

\begin{figure*}[ht]
  \centering
  \vspace{-0.15in}
  \begin{minipage}{1\linewidth}  
    \subfigure[$\varDelta f'=0~\mathrm{kHz}$, SF jammer]{
      \label{FIG.5(a)}
\includegraphics[width=0.231\linewidth,height=1.2in]{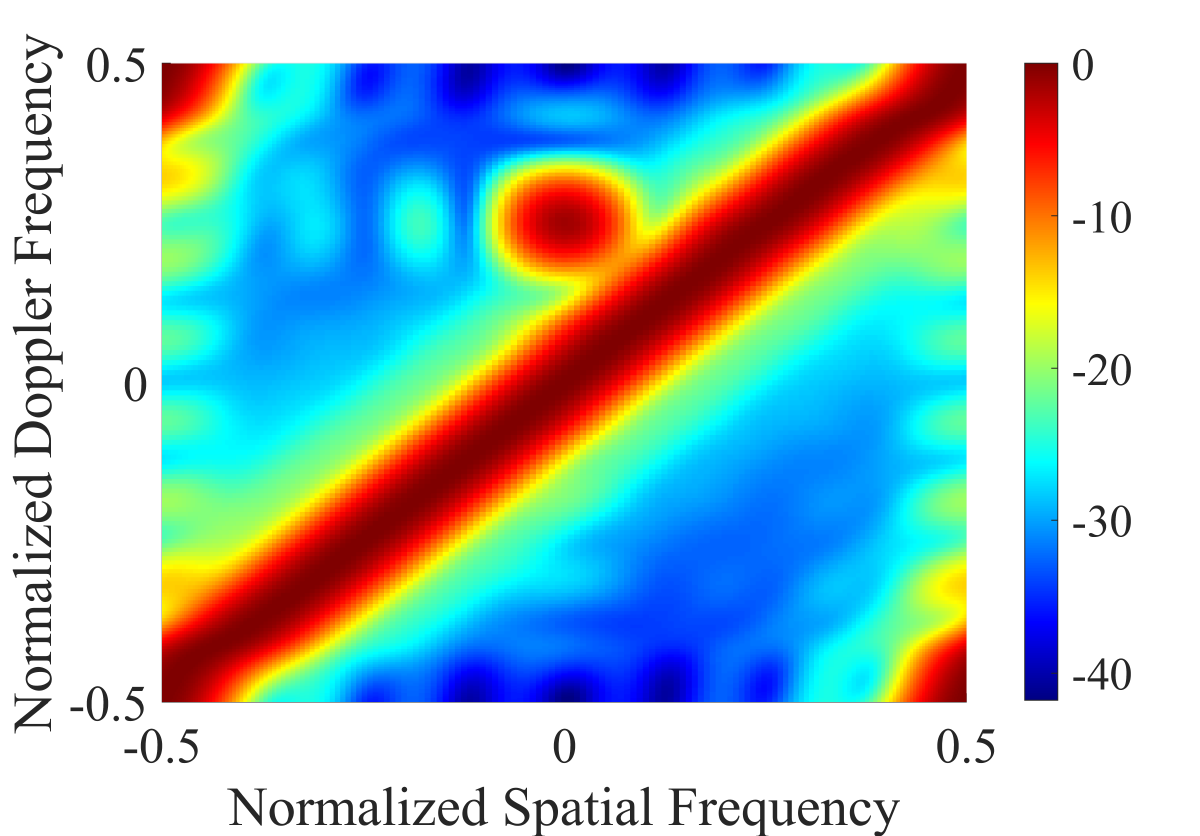}    
 }
    \subfigure[$\varDelta f'=4~\mathrm{kHz}$, SF jammer]{
      \label{FIG.5(b)}
\includegraphics[width=0.231\linewidth,height=1.2in]{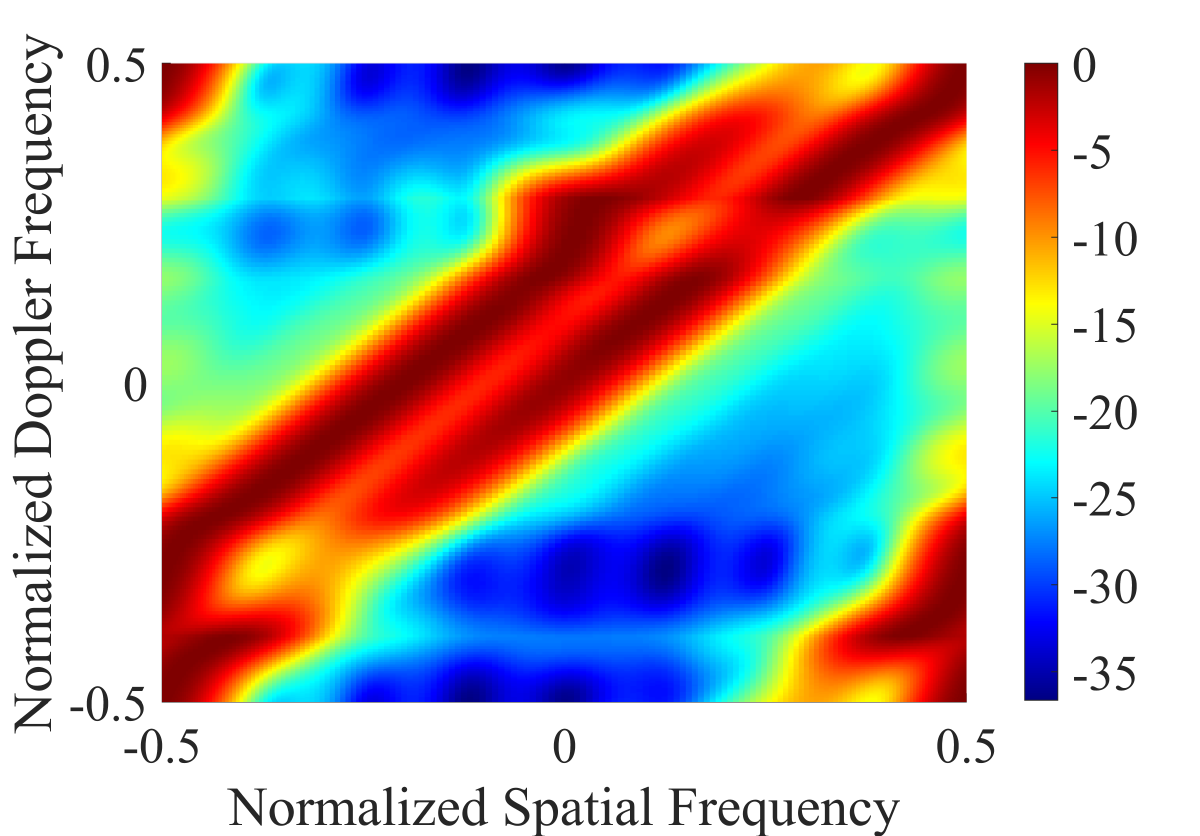}
    }
    \subfigure[$\varDelta f'=6~\mathrm{kHz}$, SF jammer]{
      \label{FIG.5(c)}
\includegraphics[width=0.231\linewidth,height=1.2in]{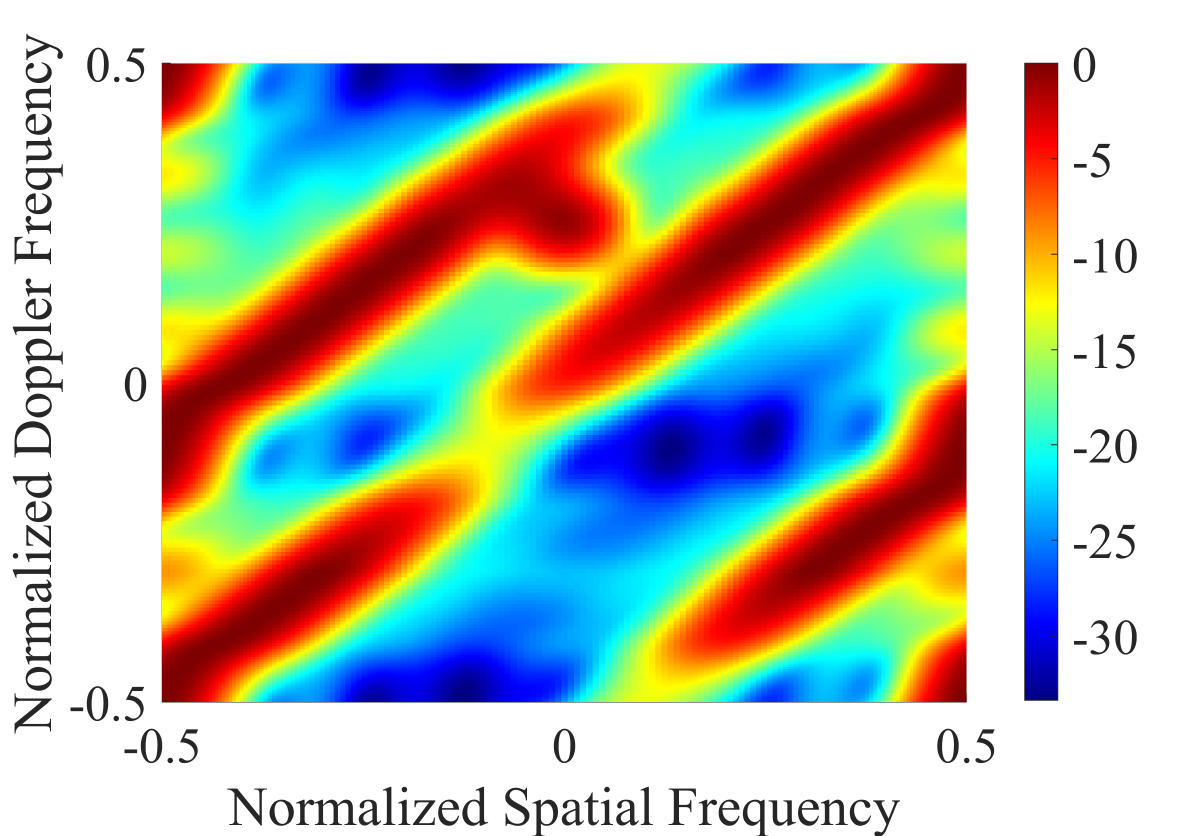}    
 }
 \subfigure[$\varDelta f'=9~\mathrm{kHz}$, SF jammer]{
      \label{FIG.5(d)}
\includegraphics[width=0.231\linewidth,height=1.2in]{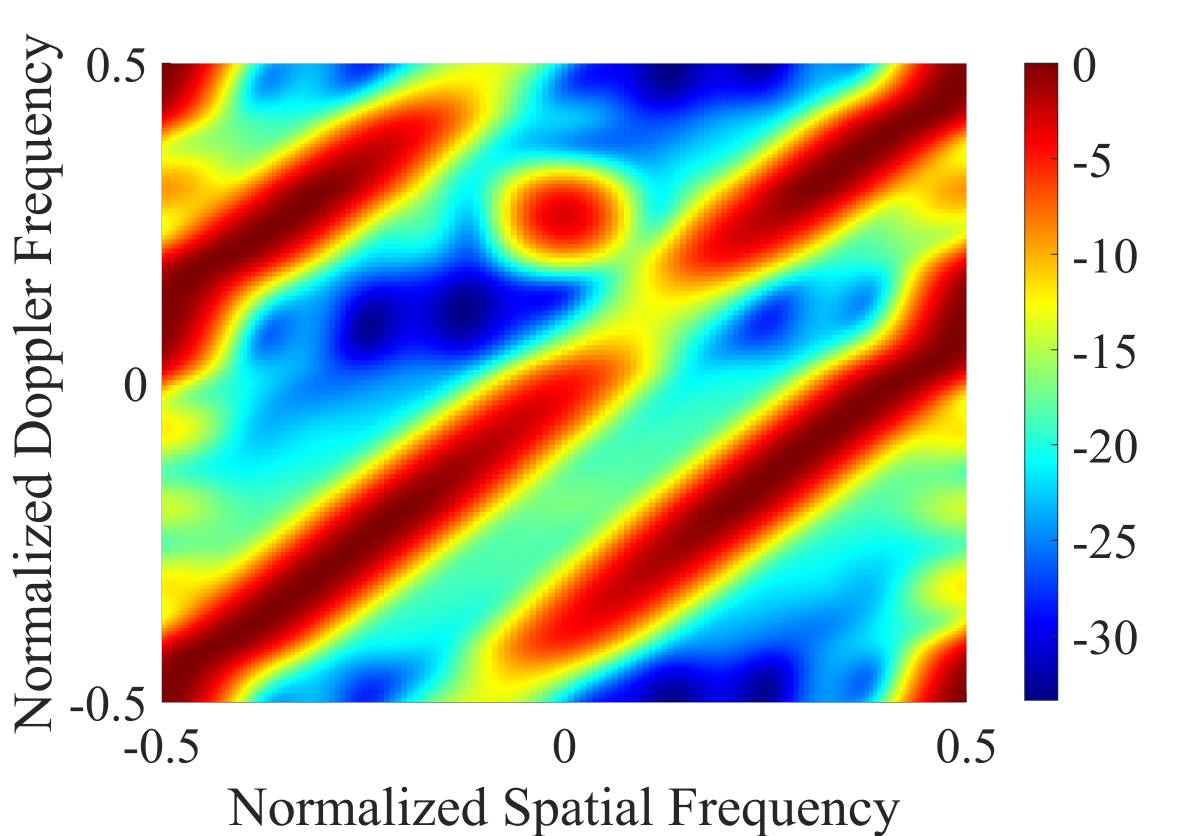}    
 }
  \end{minipage}
  \begin{minipage}{1\linewidth }
    \subfigure[$\varDelta f'=0~\mathrm{MHz}$, AF jammer]{
      \label{FIG.5(e)}
\includegraphics[width=0.231\linewidth,height=1.2in]{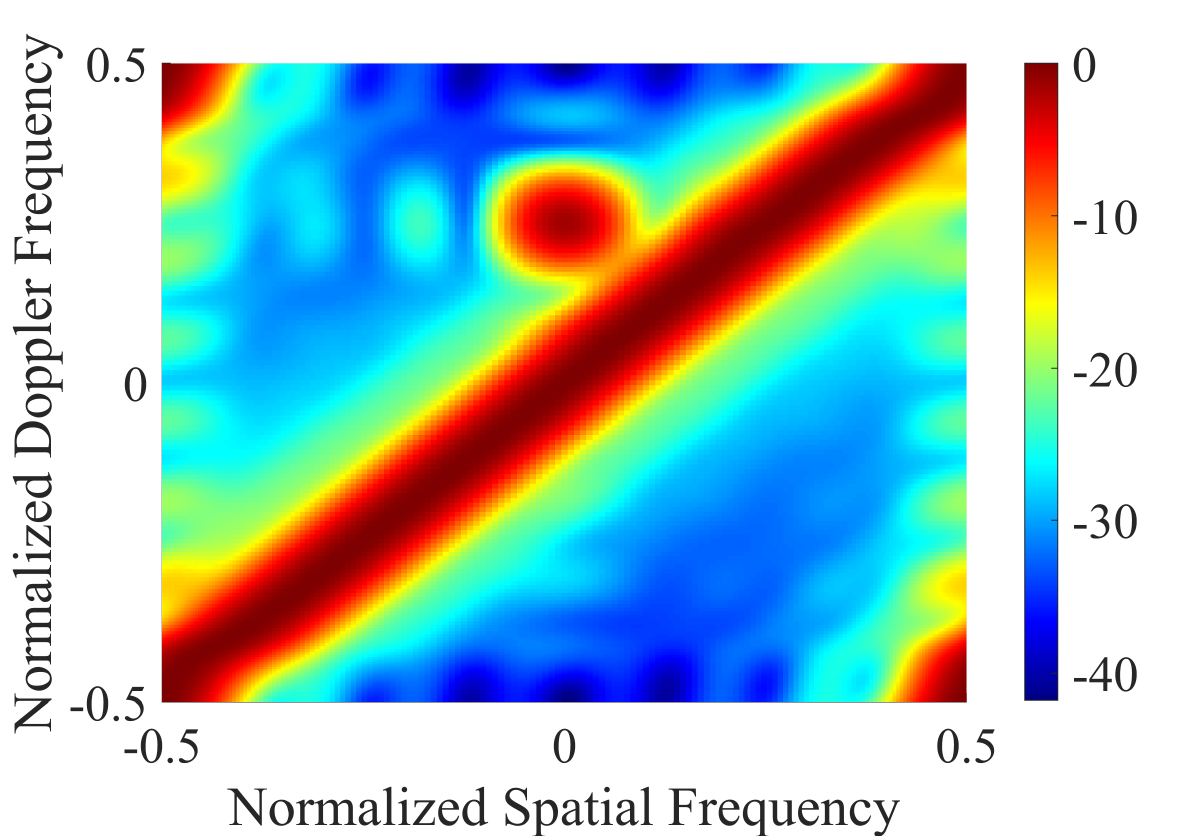}
      }
    \subfigure[$\varDelta f'=5.8~\mathrm{MHz}$, AF jammer]{\label{FIG.5(f)}  
    \includegraphics[width=0.231\linewidth,height=1.2in]{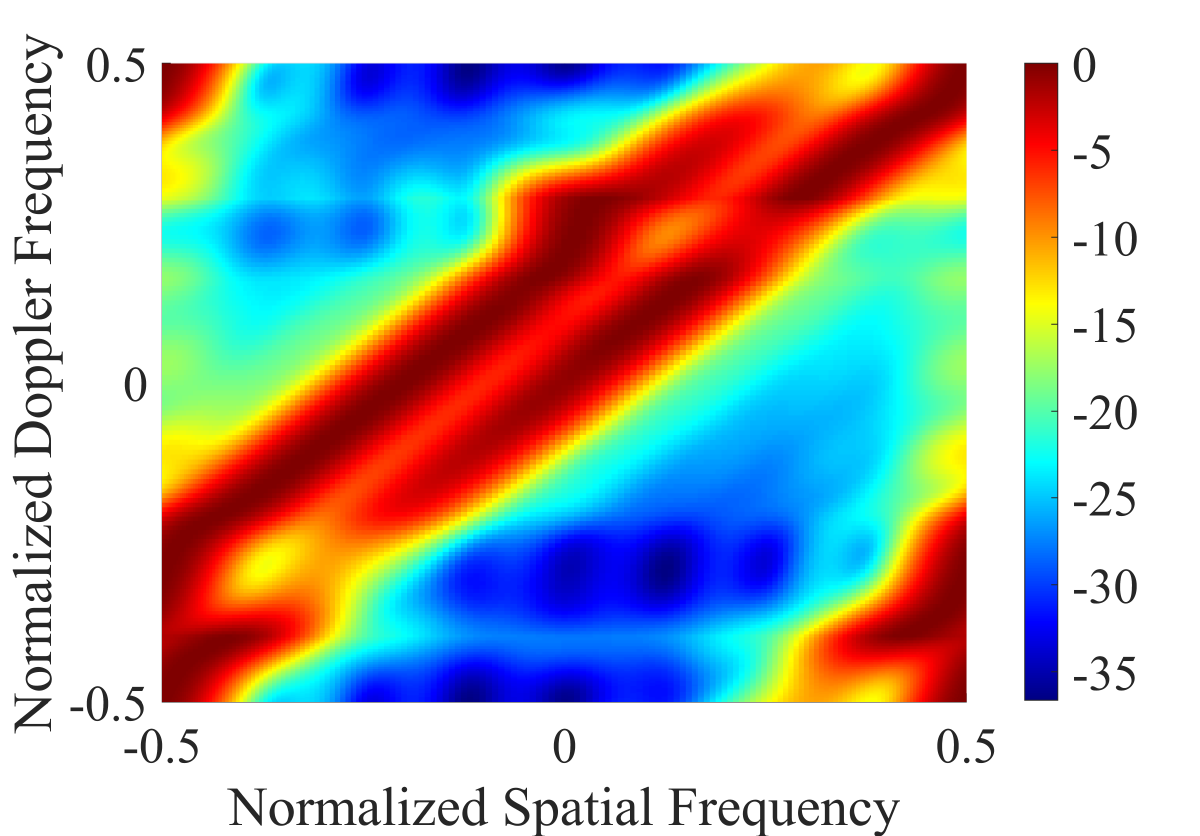}
    }
    \subfigure[$\varDelta f'=7.8~\mathrm{kHz}$, AF jammer]{
      \label{FIG.5(g)}
\includegraphics[width=0.231\linewidth,height=1.2in]{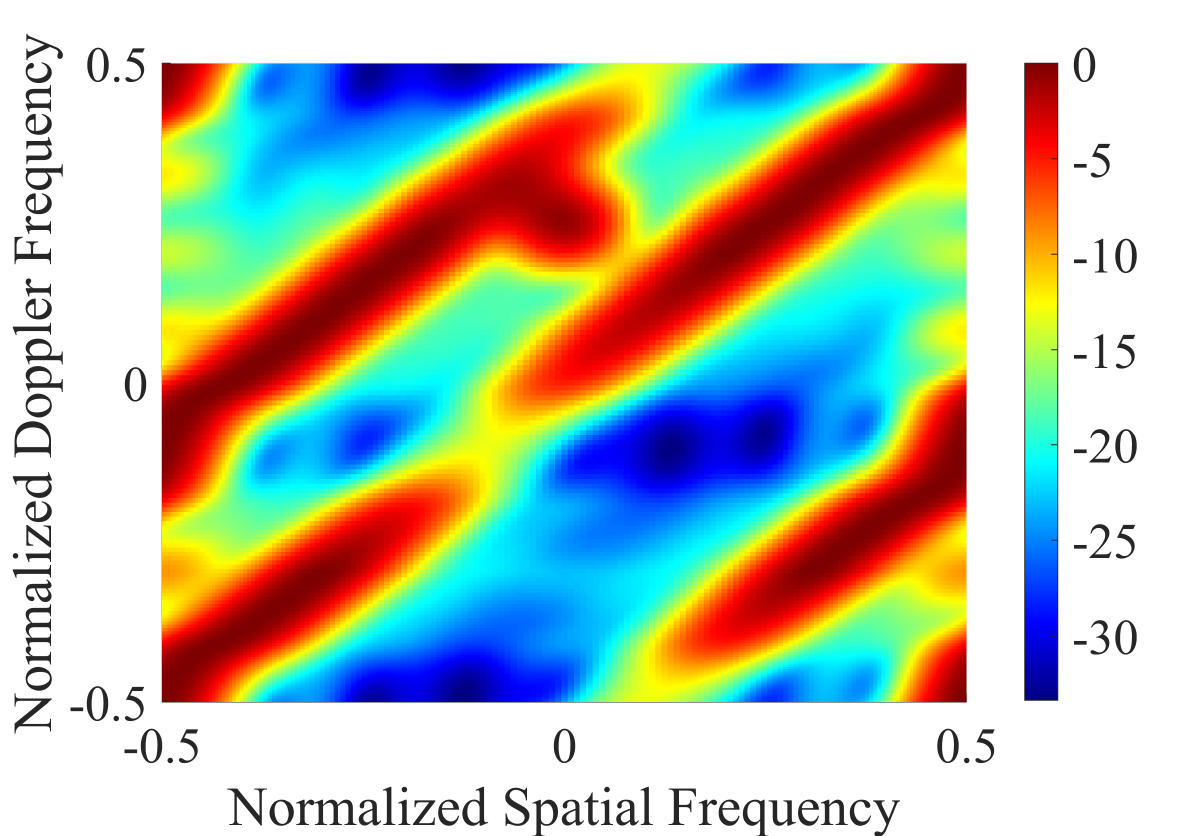}    
 }
 \subfigure[$\varDelta f'=10.8~\mathrm{kHz}$, AF jammer]{
      \label{FIG.5(h)}
\includegraphics[width=0.231\linewidth,height=1.2in]{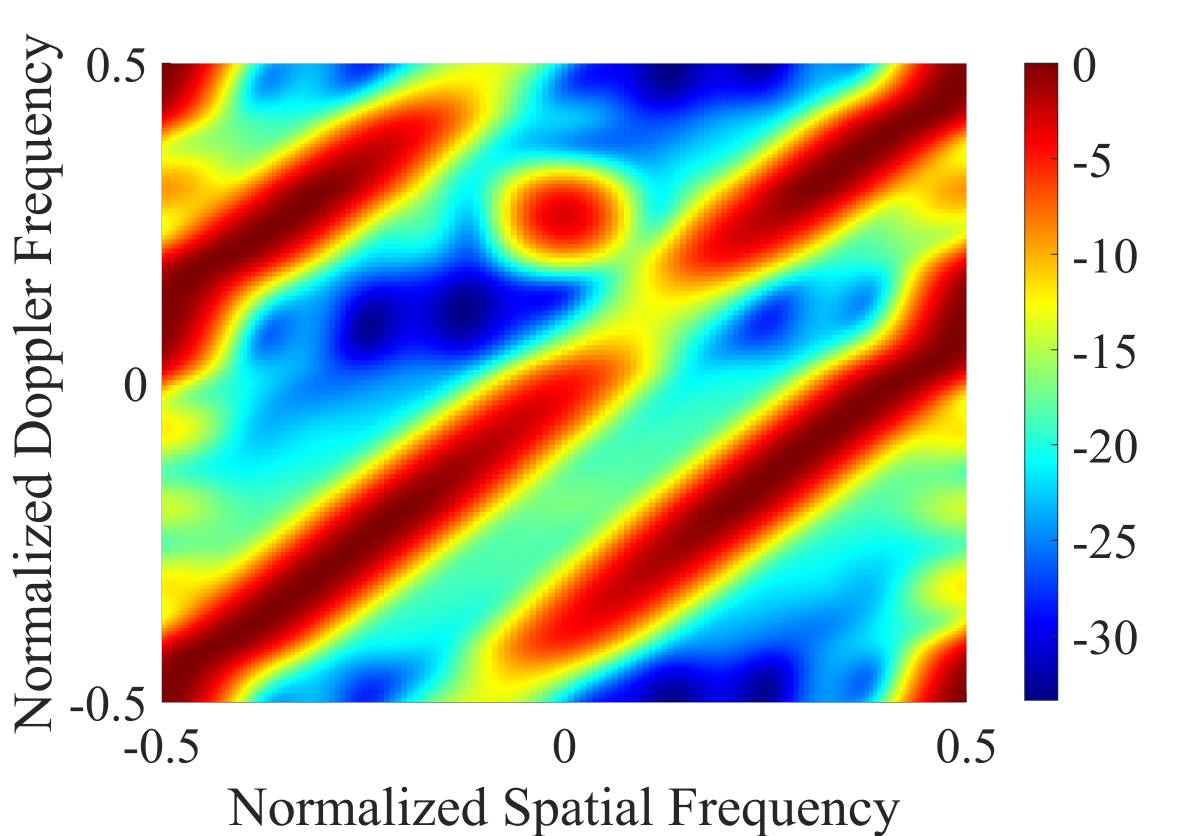}    
 }
  \end{minipage}
  \vspace{-0.10in}  
  \caption{The spatial-Doppler spectrums of four subarrays of phased-MIMO radar under different jamming frequency offsets}
  \label{FIG.5}
\end{figure*}

In Fig.\ref{FIG.3} and Fig.\ref{FIG.4}, we simulate the clutter spectrum of four subarrays of FDA-MIMO radar under the influence of SF jammers and AF jammers with different jamming frequency offsets. Both two figures demonstrate that the FDA jammer can increase the clutter rank of FDA-MIMO radar. The clutter rank ranges from approximately 19 to 37 (According to Brennan's rules [\ref{cite16}], the clutter rank of the side-looking array is proved to be 19), which strongly confirms (\ref{eq.32}) and (\ref{eq.33}). In addition, focusing on the curve corresponding to the frequency offset of 16.9 kHz, which exceeds the period of frequency offset in $\hat{\phi}_{R}^{\left( J \right)}$, it has an effect between $\Delta f'=0.1~\mathrm{kHz} $ and $\Delta f'=0.5~\mathrm{kHz} $ and confirms (\ref{eq.45a}) and (\ref{eq.45b}). Here, we use the same jamming frequency offset of FDA jamming as in Part I to illustrate that the FDA jamming can be implemented with a constant frequency offset for two functions when $\beta =1$, one is deceptive jamming discussed in Part I, the other is scattered wave jamming introduced in this part. 

\subsection{Spatial-Doppler spectrum before STAP}


In this subsection, we simulate a moving target with an azimuth of 90 degrees and a normalized Doppler frequency of 0.25. The spatial-Doppler spectrums of four subarrays phased-MIMO radar are presented in Fig.\ref{FIG.5} under the influence of two kinds of FDA jammers with different jamming frequency offsets. We can explicitly observe the effect of scattered wave FDA jamming on the power spectrum by adjusting the additional frequency offset.

In Fig.\ref{FIG.5(a)}, the clutter spectrum and jamming spectrum do not overlap with the target when $\Delta f' =0~\mathrm{KHz}$. In this case, the clutter suppression operation can be implemented through the relationship between the spatial frequency and the Doppler frequency, namely STAP, and then the target can be detected under high SCNR after clutter suppression by STAP. In Fig.\ref{FIG.5(b)} and Fig.\ref{FIG.5(c)}, the spatial-Doppler spectrum of SF jamming overlaps the target spectrum, which will seriously affect the effectiveness of phased-MIMO-STAP because suppressing the jamming will also dissipate the target energy at the same processing. The Doppler frequency shifts of $\Delta f'=4~\mathrm{kHz}$ and $\Delta f'=6$ can be calculated as $0.24$ and $0.36$, respectively. So We can see that the jamming energy covered the target energy in spatial-Doppler spectrum of Fig.\ref{FIG.5(b)} and the jamming energy departs from the target energy in spatial-Doppler spectrum of Fig.\ref{FIG.5(c)}. In Fig.\ref{FIG.5(d)}, when the frequency offset increases to $9~\mathrm{kHz}$, the spatial-Doppler spectrum of jamming shifts further without overlapping with the target. It can be concluded that the magnitude of jamming frequency offset determines the location of jamming energy in the spatial-Doppler spectrum. If the Doppler information of the moving target has been known for FDA jammer, the scattered wave FDA jamming can protect the moving target against the airborne phased-MIMO radar. For AF jamming, the required frequency offset is slightly larger than the SF jamming, which is consistent with the conclusion in (\ref{eq.42a}) and (\ref{eq.42b}). Here we use a frequency difference of 1.8 kHz to realize the case that the AF jamming signals are the same as the SF jamming signals in Fig.\ref{FIG.5(e)}, Fig.\ref{FIG.5(f)}, Fig.\ref{FIG.5(g)} and Fig.\ref{FIG.5(h)}. These simulations illustrate the correctness of the frequency offset discussion in section V.

\begin{figure}[t]
\centerline{\includegraphics[width=20pc]{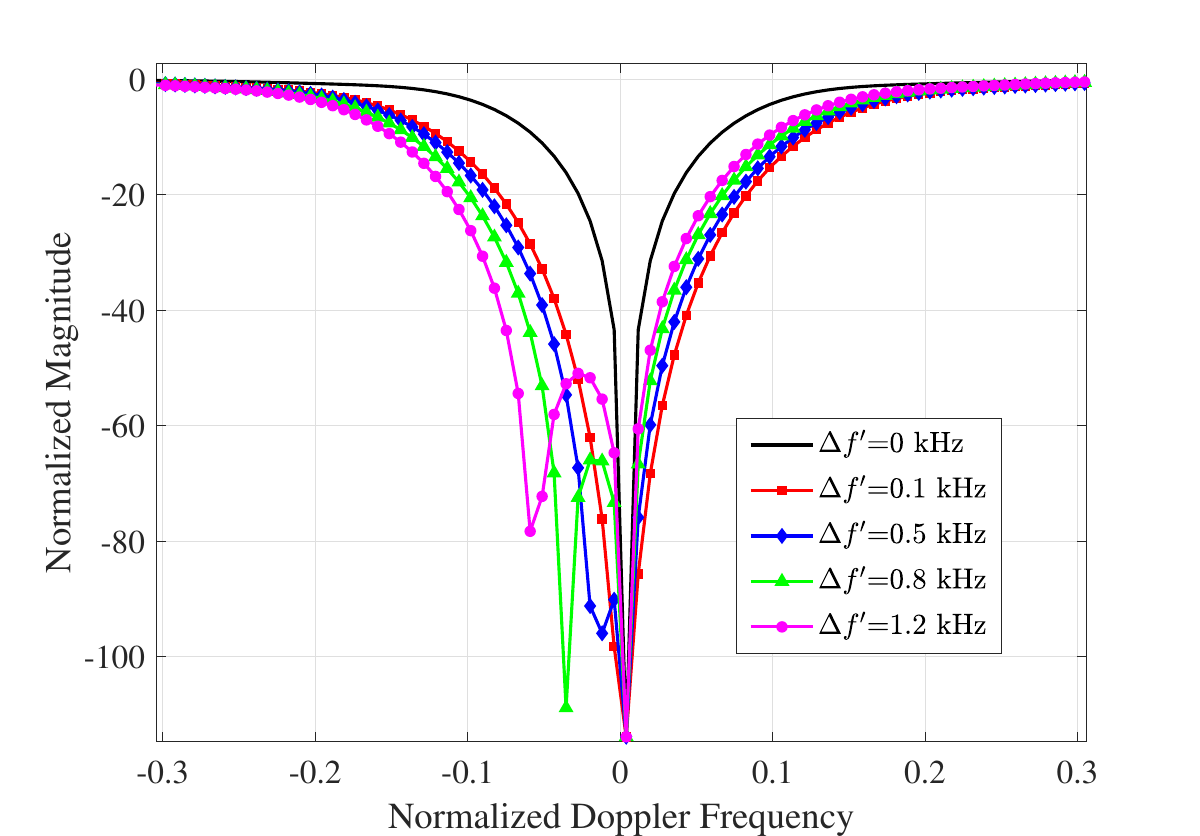}}
\caption{Effect of the small $\varDelta f'$ for SF jamming.}
\label{FIG.6}
\end{figure}

 
\subsection{IF of phased-MIMO-STAP}

\begin{figure}[t]
\centerline{\includegraphics[width=20pc]{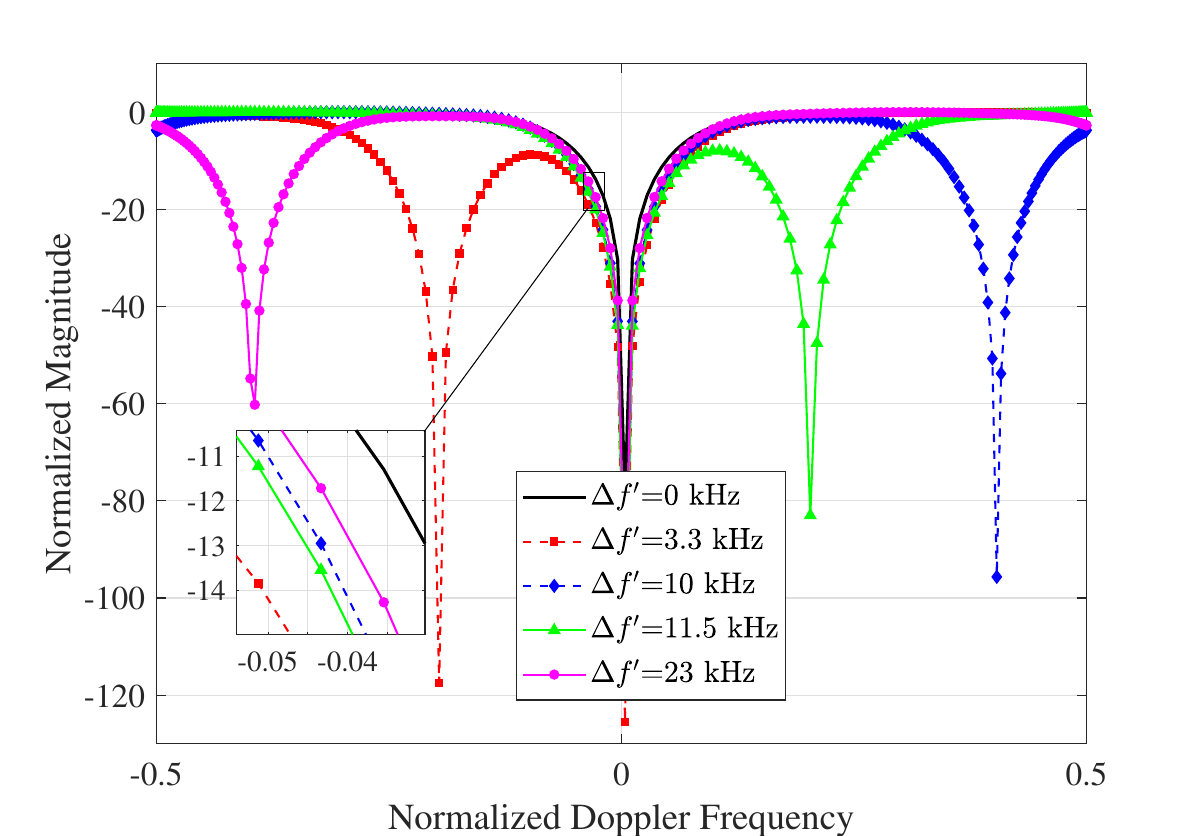}}
\caption{Effect of the big $\varDelta f'$ for SF jamming.}
\label{FIG.7}
\end{figure}

\begin{figure}[t]
\centerline{\includegraphics[width=20pc]{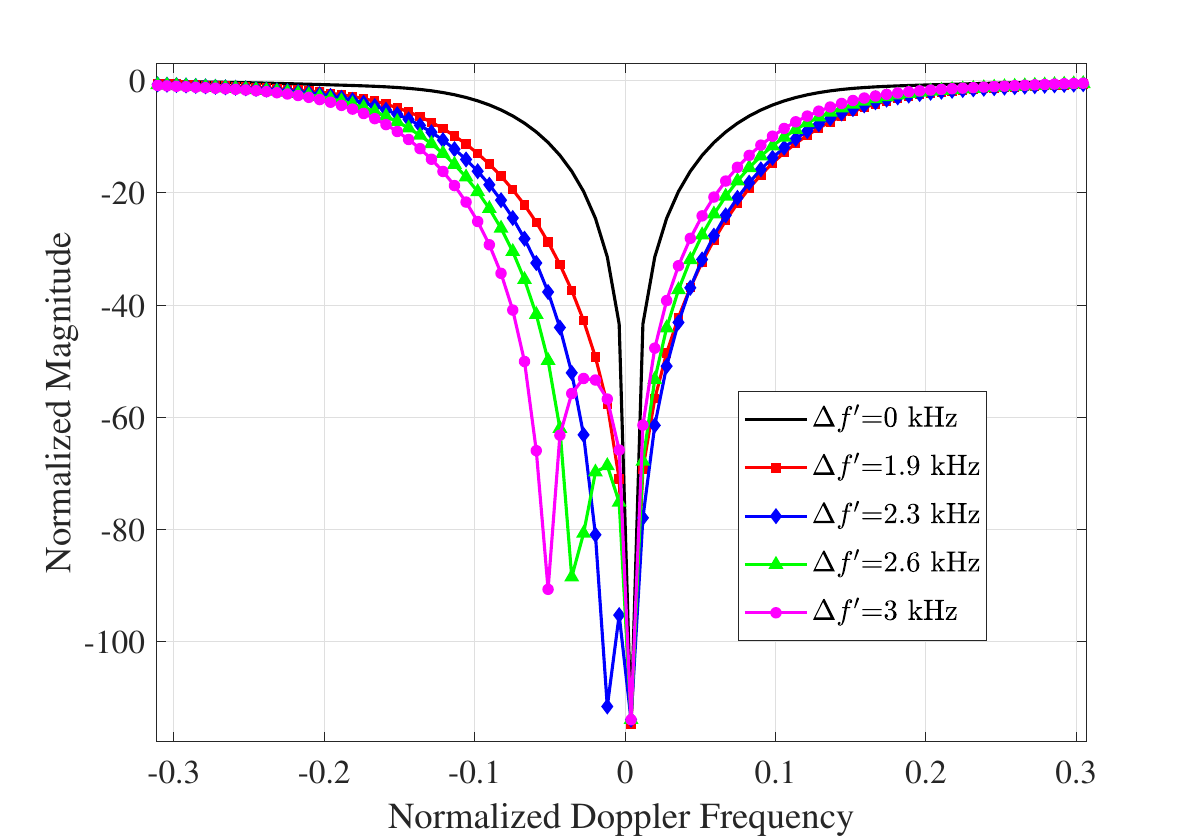}}
\caption{Effect of the small $\varDelta f'$ for AF jamming.}
\label{FIG.8}
\end{figure}

\begin{figure}[t]
\centerline{\includegraphics[width=20pc]{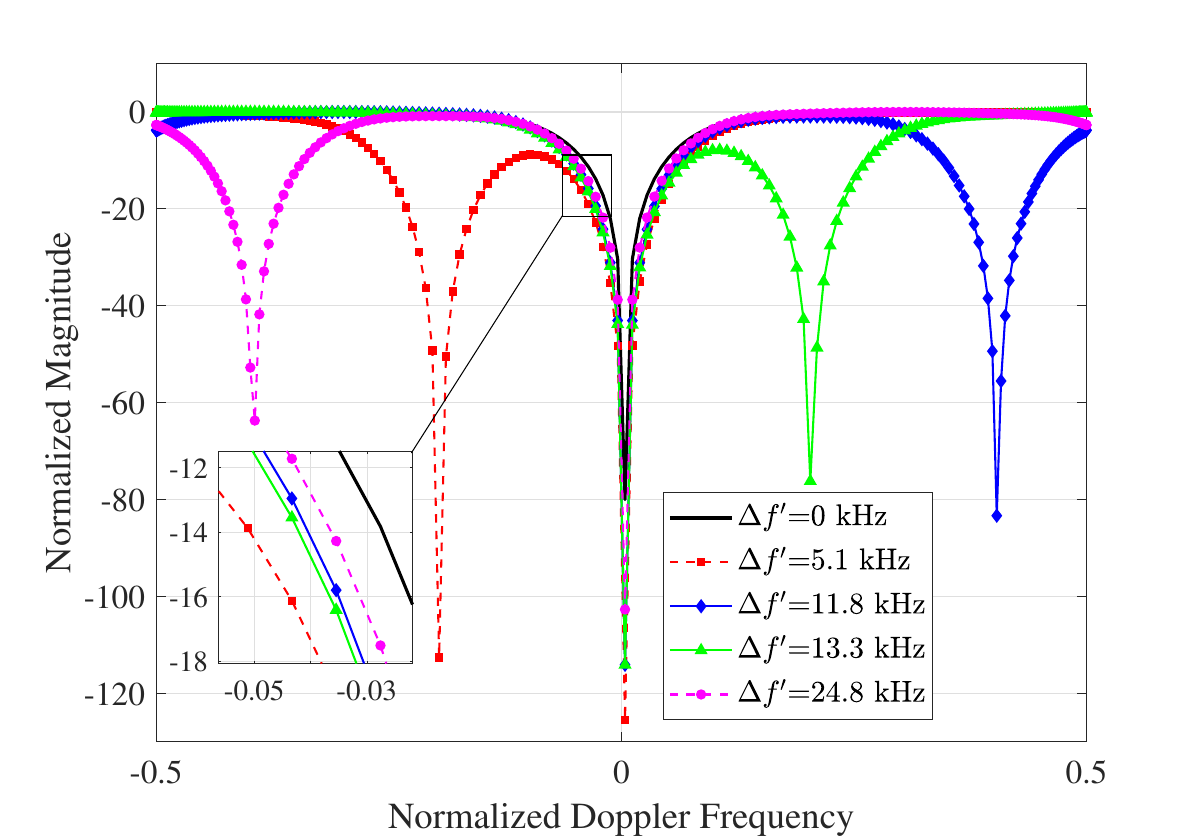}}
\caption{Effect of the big $\varDelta f'$ for AF jamming.}
\label{FIG.9}
\end{figure}

In this subsection, we simulate the IF of phased-MIMO-STAP to evaluate the performance of FDA jamming signals with different frequency offsets. We classify the jamming frequency offset into a large group and a small group in order to explicitly simulate the effect of the jamming frequency offset on the IF, as it is an important measure for the performance of phased-MIMO-STAP clutter suppression. The SF jammer with small jamming frequency offset and big jamming frequency offset have been presented in Fig.\ref{FIG.6} and Fig.\ref{FIG.7}, respectively. The AF jammer with small jamming frequency offset and big jamming frequency offset have been presented in Fig.\ref{FIG.8} and Fig.\ref{FIG.9}, respectively. Each numerical result is finished with 100 Monte Carlo simulations. Noted that $\varDelta f'=0~\mathrm{kHz}$ represents the conventional scattered wave jamming proposed in [\ref{cite1}] as a comparison. 

In Fig.\ref{FIG.6}, the IF of phased-MIMO-STAP is shown under the conditions of SF jammers with a small frequency offset. Compared with the curve of conventional scattered wave jamming [\ref{cite18}], the SF jammer with a small frequency offset can widen the notch of IF, which directly increases the minimum detectable velocity (MDV) of airborne radar [\ref{cite2}]. As the jamming frequency offset grows up, the second If notch gradually appears to the left of the 0 Doppler frequency, which is consistent with the conclusions of (\ref{eq.42a}).

In Fig.\ref{FIG.7}, the IF of phased-MIMO-STAP is shown under the conditions of SF jammer with a big frequency offset. As the frequency offset increases, we can see two consequences, one is that the second notch shifts as the variation of jamming frequency offset, and the other is that the second notch becomes shallower as the frequency offset increases. The increasing of frequency offset directly causes the energy loss of the scattered wave jamming signal after MF. Through the calculation, the Doppler position of the second notch is verified the correctness of (\ref{eq.42a}). From the zoomed-in perspective of the main notch (Doppler frequency is 0), the influence of SF jammer with the larger frequency offset is weakened for the main notch of IF, because the difference between the SF interference signal and the clutter signal becomes larger.

In Fig.\ref{FIG.8} and Fig.\ref{FIG.9}, we also simulated the comparative results for AF jammer against phased-MIMO-STAP, which can be indicated the same conclusion as SF jammers in Fig.\ref{FIG.6} and Fig.\ref{FIG.7}.


\begin{table*}[!b]\centering
 \caption{Comparison of the FDA jamming and existing Doppler jamming techniques}
 \label{table.2}
 \renewcommand\arraystretch{1.9}
 \begin{minipage}{\textwidth}\centering
 \scalebox{0.85}{\begin{tabular}{cccccccc}
 \hline
 \textbf{Jamming Technique}  &  \textbf{Counter Objective}  &  \textbf{Function}   &  \textbf{Results} &  \textbf{Drawbacks} \\
 \hline
\multirow{2}{*}{\makecell{FDA-SW jamming\\(Against phased-MIMO-STAP)}} &\multirow{2}{*}{Radar detection} & \multirow{2}{*}{Reduce output SCNR}     & \multirow{2}{*}{Shift the IF notch} & \multirow{2}{*}{\makecell{Sophisticated jamming antennas\\target information}}\\
      & &     &  &  \\                               
\hline
\multirow{3}{*}{\makecell{Doppler deceptive jamming\textsuperscript{[\ref{cite13}-\ref{cite15}]}\\ (No scene constraints)}} &  Radar detection\textsuperscript{[\ref{cite13}]}  & Reduce detection probability    & False target information & \multirow{3}{*}{\makecell{Widely and systematically studied\\Easily suppressed and recognized}} \\            
             &  Radar tracking\textsuperscript{[\ref{cite14}]} & Reduce tracking accuracy   & False trajectory &    \\
             &  Radar imaging\textsuperscript{[\ref{cite15}]} & Disrupte imaging   & Interfere Doppler compensation &    \\
\hline
\multirow{2}{*}{\makecell{Doppler towing jamming\\ (Missiles or vessels)}} & Radar detection\textsuperscript{[\ref{cite16}]} &  Fake Doppler information   &  Cover decoy & Desirable hardware materials  \\
             &  Radar tracking\textsuperscript{[\ref{cite17}]} & Fake wave gate    & False target decoy   &   Complex trajectory deception  \\
 \hline
 \multirow{2}{*}{\makecell{Scattered wave jamming\textsuperscript{[\ref{cite18}]}\\ (Against SAR or STAP radar)}} & \multirow{2}{*}{Radar imaging} &  False real-time location &  \multirow{2}{*}{Increase false targets}  &\multirow{2}{*}{\makecell{Require more prior information\\ and computationally complex}}   \\
             &    & False Doppler information  &  &    \\
 \hline
 \end{tabular}}
 \end{minipage}
 \vspace{0mm}
 \end{table*}

\section{DISCUSSION AND PROSPECT}

In this section, we first discuss the difference between the proposed FDA-SW jamming and the existing Doppler jamming techniques. The existing Doppler jammers are mainly categorized into Doppler deceptive jamming, Doppler towing jamming and scattered wave jamming. Doppler deceptive jamming provides false Doppler information to radar to interfere  detection and tracking [\ref{cite13}]-[\ref{cite15}]. However, we don't find a basis for Doppler deceptive jamming against phased-MIMO radar from the existing references. Doppler towing jamming primarily interferes with radar target tracking and doesn't involve airborne STAP [\ref{cite16}]-[\ref{cite17}]. Scattered wave jamming is mainly proposed against SAR radar in the existing references [\ref{cite18}]. Since their application scenarios are not consistent with the proposed jamming in this paper, we cannot compare all of them in the same scenario against phased-MIMO-STAP. Nevertheless, we concluded their counter objectives, functions, results, and drawbacks in Table \ref{table.2}. We expect that such a summary can provide inspiration for subsequent research on FDA jamming.

As a novel framework for ECM, our works in two-part papers of this series have some limitations. This section lists the valuable prospects of FDA jamming, which are expected to be investigated by intrigued researchers. 
\begin{itemize}
  \item [1)] 
  The categories and application scenarios of FDA jammers are expected to be expanded. The combination of FDA and other hot jamming technologies can be thoroughly investigated, such as interrupted sampling repeater jamming [\ref{cite26}, \ref{cite27}], range-Doppler deceptive jamming [\ref{cite28}], and towed jamming [\ref{cite29}]. Accordingly, the performance evaluation and frequency offset selection of FDA jammers in different application scenarios must be further discussed. 
  \item [2)]
  The effects of FDA jammers on parameter estimation, target tracking and surveillance process are expected to be investigated. In fact, the phase shift caused by jamming frequency offset directly deteriorates the covariance of the sample data, which causes modifications in the statistical properties of the observed data [\ref{cite30}]. For example, the frequency offset of FDA jamming affects the clutter subspace, so the performance loss of subspace-based estimation algorithms needs to be evaluated.
  \item [3)]
  The quantitative relationship between jamming signal power and frequency offset is expected to be systematically investigated [\ref{cite31}]. Although the jammer transmitter power is typically greater than the radar transmitter power, the jamming energy loss caused by the mismatch of jamming frequency offset cannot neglected easily.
  \item [4)]
  The effective suppression methods against FDA jamming are expected to be investigated. The compensation of frequency offset may be effective in equalizing the effects of FDA jamming, but the identification of FDA jamming signals needs to be completed in advance [\ref{cite32}]. Thanks to the development of machine learning in signal processing, the identification of FDA interference may be accomplished by utilizing spectral images [\ref{cite33}].
\end{itemize}

\section{CONCLUSION}

This paper has documented the effectiveness of the FDA-SW jamming against phased-MIMO-STAP. Firstly, the scatterer trajectory equation of FDA-SW jamming which can determine the spatial frequency and Doppler frequency of jamming signals, has been derived. Next, we prove that the FDA-SW jamming could effectively change the clutter rank and deteriorate STAP performance. All theoretical analysis are developed for SF and AF jammers against two cases of phased-MIMO radar. The numerical results have verified the theoretical analyses and illustrated the well-function of FDA-SW jamming.

In summary, this two-part series of papers proposed the FDA jamming inspired by the advantages of FDA radar and analyzed the effectiveness of FDA jamming in countering the airborne phased-MIMO radar. Firstly, we categorize FDA jammers as SF and AF jammers based on their emission mechanism. According to the derivations of output matched-filtering, FDA jamming can be used as the support jammer when the jammer is far away from the target. A single FDA jammer can generate multiple range-dimensional false targets after the MF process of the radar receiver. Through the establishment of three measures for spatial filtering, FDA jamming also can be used as the self-protection jammer when the jammer is close to the target in the range dimension or azimuth dimension, namely sidelobe or mainlobe jamming. From the perspective of countering phased-MIMO-STAP, we proposed the FDA-SW jamming to deteriorate the clutter suppression and protect the moving target. On the one hand, FDA-SW jamming can increase the clutter rank of phased-MIMO with more than one subarray. On the other hand, the FDA-SW jamming can shift the IF notch of phased-MIMO-STAP by adjusting the jamming frequency offset, which can worsen the MDV and protect the target. All numerical results have verified the theoretical analysis and effectiveness of the proposed FDA jamming. We expect to further research of novel frameworks on the combination of FDA jamming and other outstanding jamming techniques.


\begin{figure}[t]
\centerline{\includegraphics[width=19pc]{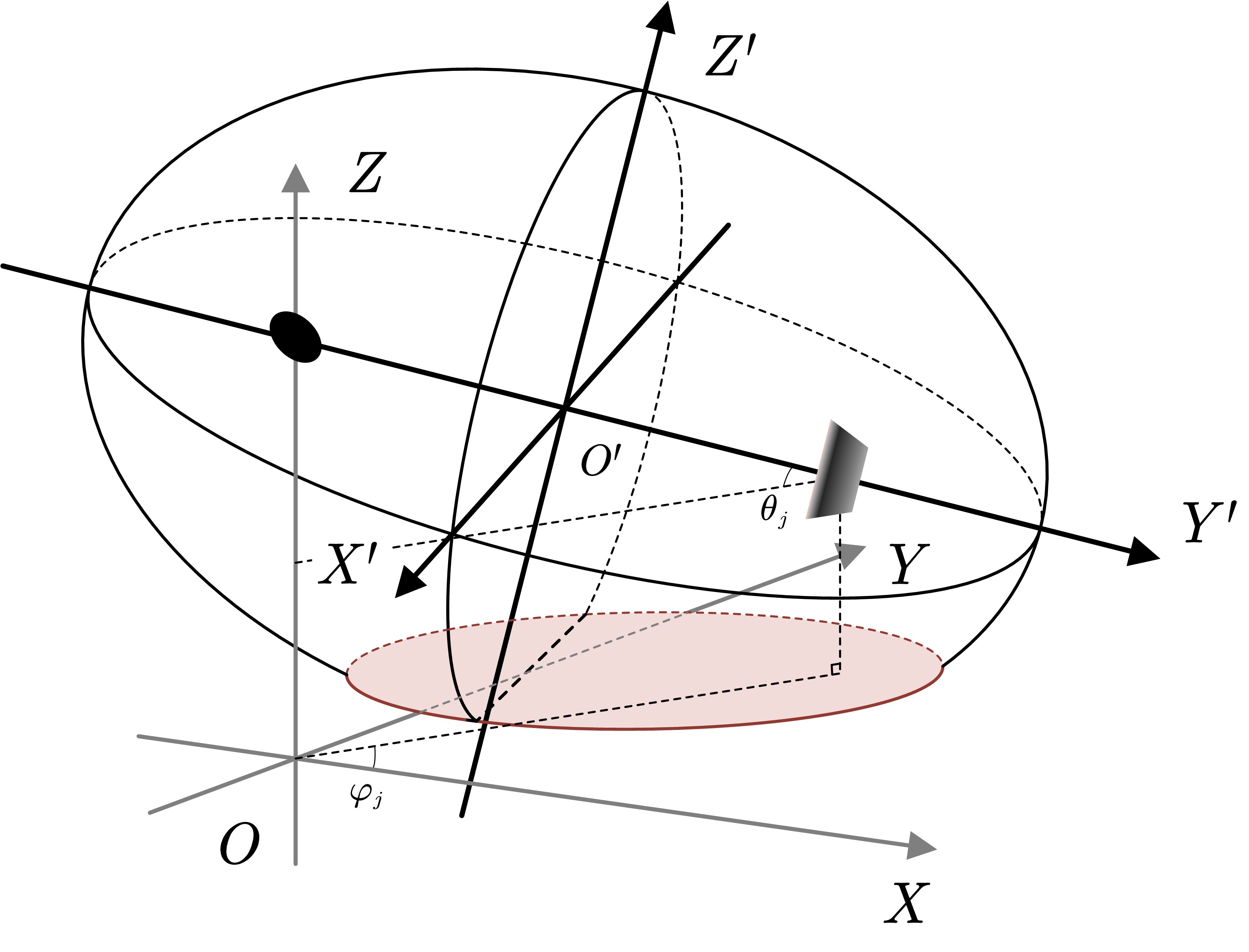}}
\caption{The trajectory of the ground scatterers for FDA scattered wave jamming.}
\label{FIG.A}
\end{figure}

\setcounter{TempEqCnt}{\value{equation}} 
\setcounter{equation}{46}
\begin{figure*}[t]
\begin{align}
\left[ \begin{array}{c}
  x'\\
  y'\\
  z'\\
\end{array} \right] =\left[ \begin{matrix}
  \cos \left( \frac{\pi}{2}-\varphi _j \right)&   \sin \left( \frac{\pi}{2}-\varphi _j \right)&   0\\
  -\sin \left( \frac{\pi}{2}-\varphi _j \right)&    \cos \left( \frac{\pi}{2}-\varphi _j \right)&   0\\
  0&    0&    1\\
\end{matrix} \right] \left[ \begin{matrix}
  \cos \theta _j&   0&    \sin \theta _j\\
  0&    1&    0\\
  -\sin \theta _j&    0&    \cos \theta _j\\
\end{matrix} \right] \left[ \begin{array}{c}
  x\\
  y\\
  z\\
\end{array} \right] +\left[ \begin{array}{c}
  \frac{R_f}{2}\cos \varphi _j\cos \theta _j\\
  \frac{R_f}{2}\sin \varphi _j\cos \theta _j\\
  H-\frac{R_f}{2}\sin \theta _j\\
\end{array} \right] 
\label{eq.A2}
\end{align}
\hrulefill
\end{figure*}
\setcounter{equation}{\value{TempEqCnt}} 
\setcounter{equation}{45}

\section*{Appendix A}
\label{Appendix.A}
In order to establish the standard elliptic equation, we construct the local coordinate system $X'Y'Z'$, as shown in Fig.\ref{FIG.A}, which is transformed by a clockwise rotation of the global coordinate system around the Z axis by $\pi/2-\varphi_j$ and the X axis by $\theta_j$, and three translations along the coordinate axes. In the local coordinate system, the standard elliptic equation can be written as

\begin{equation}
\frac{x'^2}{R_t^{2}}+\frac{y'^2}{R_t^{2}-\left( R_f/2 \right) ^2}+\frac{z'^2}{R_t^{2}-\left( R_f/2 \right) ^2}=1
\label{eq.A1}
\end{equation}
The relationship between the global and local coordinate systems can be expressed by the following equation. After solving the equation in (\ref{eq.A2}), then
\setcounter{TempEqCnt}{\value{equation}} 
\setcounter{equation}{47}
\begin{equation}
\begin{cases}
  x'=&x\sin \varphi _j\cos \theta _j+y\cos \varphi _j+z\sin \varphi _j\sin \theta _j\\
  &+\frac{R_f}{2}\cos \varphi _j\cos \theta _j\\
  y'=&-x\cos \varphi _j\cos \theta _j+y\sin \varphi _j-z\cos \varphi _j\sin \theta _j\\
  &+\frac{R_f}{2}\sin \varphi _j\cos \theta _j\\
  z'=&H-\left( x+\frac{R_f}{2} \right) \sin \theta _j+z\cos \theta _j\\
\end{cases}
\label{eq.A3}
\end{equation}
Assume that the global coordinates of the ground scattering point for jamming is $\hat{\boldsymbol{p}}_i=\left( \hat{x}_i,\hat{y}_i,0 \right)$. Substituting the coordinates value into (\ref{eq.A3}) and (\ref{eq.A2}), the trajectory equation can be written as 
\begin{align}
&\frac{\left( \hat{x}_i\sin \varphi _j\cos \theta _j+\hat{y}_i\cos \varphi _j+\frac{R_f}{2}\cos \varphi _j\cos \theta _j \right) ^2}{R_{t}^{2}}
\nonumber\\
+&\frac{\left( -\hat{x}_i\cos \varphi _j\cos \theta _j+\hat{y}_i\sin \varphi _j+\frac{R_f}{2}\sin \varphi _j\cos \theta _j \right) ^2}{R_{t}^{2}-\left( R_f/2 \right) ^2}
\nonumber\\&~~~~~~~~~~~+\frac{\left[ H-\left( \hat{x}_i+\frac{R_f}{2} \right) \sin \theta _j \right] ^2}{R_{t}^{2}-\left( R_f/2 \right) ^2}=1
\label{eq.A4}
\end{align}
Expanding (1) to a general elliptic equation in the global coordinate system, then
\begin{equation}
Q_1\hat{x}_{i}^{2}+Q_2\hat{x}_i\hat{y}_i+Q_3\hat{y}_{i}^{2}+Q_4\hat{x}_i+Q_5\hat{y}_i+Q_6=0
\label{eq.A5}
\end{equation}
where
\begin{subequations}
\begin{align}
Q_1&=\left( \frac{\sin \varphi _j\cos \theta _j}{R_t} \right) ^2+\frac{\left( \cos \varphi _j\cos \theta _j \right) ^2+\left( \sin \theta _j \right) ^2}{R_{t}^{2}-\left( R_f/2 \right) ^2}\label{eq.A6a}
\\
Q_2&=\frac{2\cos \varphi _j\sin \varphi _j\cos \theta _j}{R_{t}^{2}}-\frac{2\cos \varphi _j\sin \varphi _j\cos \theta _j}{R_{t}^{2}-\left( R_f/2 \right) ^2}\label{eq.A6b}
\\
Q_3&=\left( \frac{\cos \varphi _j}{R_t} \right) ^2+\frac{\left( \sin \varphi _j \right) ^2}{R_{t}^{2}-\left( R_f/2 \right) ^2}\label{eq.A6c}
\\
Q_4&=\frac{R_f\cos \varphi _j\sin \varphi _j\left( \cos \theta _j \right) ^2}{2R_{t}^{2}}+\frac{R_f\left( \sin \theta _j \right) ^2}{R_{t}^{2}-\left( R_f/2 \right) ^2}\nonumber\\&-\frac{2H\sin \theta _j}{R_{t}^{2}-\left( R_f/2 \right) ^2}-\frac{R_f\cos \varphi _j\sin \varphi _j\left( \cos \theta _j \right) ^2}{2\left[ R_{t}^{2}-\left( R_f/2 \right) ^2 \right]}\label{eq.A6d}
\\
Q_5&=-\frac{R_f\left( \cos \varphi _j \right) ^2\cos \theta _j}{2R_{t}^{2}}+\frac{R_f\left( \sin \varphi _j \right) ^2\cos \theta _j}{2\left[ R_{t}^{2}-\left( R_f/2 \right) ^2 \right]}\label{eq.A6e}
\\
Q_6&=\left( \frac{R_f\cos \varphi _j\cos \theta _j}{2R_t} \right) ^2+\frac{\left( R_f\sin \varphi _j\cos \theta _j \right) ^2}{\left( 2R_t \right) ^2-\left( R_f \right) ^2}\nonumber\\&+\frac{\left( H-R_f\sin \theta _j/2 \right) ^2}{R_{t}^{2}-\left( R_f/2 \right) ^2}\label{eq.A6f}
\end{align}
\end{subequations}
Explicitly, this is an ellipse in the XOY plane and its standard equation can be written as
\begin{equation}
\frac{\left( \hat{x}_i-F_x \right) ^2}{L_{a}^{2}}+\frac{\left( \hat{y}_i-F_y \right) ^2}{L_{b}^{2}}=1
\label{eq.A7}
\end{equation}
where the coordinate of the ellipse centre is $\left( F_x,F_y,0 \right)$, the long and short axles are $L_a$ and $L_b$. 
\begin{subequations}
\begin{align}
F_x=&\frac{Q_2Q_5-2Q_3Q_4}{4Q_1Q_3-Q_{2}^{2}}\label{eq.A8a}
\\
F_y=&\frac{Q_2Q_4-2Q_1Q_5}{4Q_1Q_3-Q_{2}^{2}}\label{eq.A8b}
\\
L_a=&\sqrt{\frac{2\left( Q_1x_{0}^{2}+Q_3y_{0}^{2}+Q_2x_0y_0-1 \right)}{Q_1+Q_3+\sqrt{\left( Q_1-Q_3 \right) ^2+Q_{2}^{2}}}}\label{eq.A8c}
\\
L_b=&\sqrt{\frac{2\left( Q_1x_{0}^{2}+Q_3y_{0}^{2}+Q_2x_0y_0-1 \right)}{Q_1+Q_3-\sqrt{\left( Q_1-Q_3 \right) ^2+Q_{2}^{2}}}}\label{eq.A8d}
\end{align}
\end{subequations}

\section*{APPENDIX B}

Taking the SF jammer as an example for derivation, the AF jammer can be straightforward derived. Let
\begin{equation}
H\left( t \right) =\mathrm{Rect}\left( \frac{t-\frac{T_p}{2}}{T_p} \right) \frac{\sin P\pi \left( \varDelta f't-\hat{\phi}_{R}^{\left( J \right)} \right)}{\sin \pi \left( \varDelta f't-\hat{\phi}_{R}^{\left( J \right)} \right)}
\label{eq.B1}
\end{equation}
where
\begin{equation}
\mathrm{Rect}\left( \frac{t-\frac{T_p}{2}}{T_p} \right) =\begin{cases}
  1, t\leqslant \left| \frac{T_p}{2} \right|\\
  0, \mathrm{otherwise}\\
\end{cases}
\label{eq.B2}
\end{equation}
Its Fourier transform can be expressed as 
\begin{equation}
\mathcal{F} \left\{ \mathrm{Rect}\left( \frac{t-\frac{T_p}{2}}{T_p} \right) \right\} =e^{-j\pi fT_p}T_p\sin\mathrm{c}\left( fT_p \right) 
\label{eq.B3}
\end{equation}
where
\begin{equation}
\sin\mathrm{c}\left( fT_p \right) =\frac{\sin \left( \pi fT_p \right)}{\pi fT_p}
\label{eq.B4}
\end{equation}
Then, the Fourier transform of $H\left( t \right)$ can be expressed in (\ref{eq.B5}), 
\begin{align}
&\mathcal{F} \left\{ H\left( t \right) \right\} 
\nonumber\\
=&\frac{1}{2\mathrm{\pi}}\mathcal{F} \left\{ \mathrm{Rect}\left( \frac{t-\frac{T_p}{2}}{T_p} \right) \right\} \circledast \mathcal{F} \left\{ \frac{\sin P\pi \left( \varDelta f't-\hat{\phi}_{R}^{\left( J \right)} \right)}{\sin \pi \left( \varDelta f't-\hat{\phi}_{R}^{\left( J \right)} \right)} \right\} 
\nonumber\\
=&T_pe^{j\mathrm{\pi}\left( P-1 \right) \phi _{R}^{\left( J \right)}}
\nonumber\\
 &\times \sum_{i=1}^P{e^{-j\mathrm{\pi}f\left( \frac{4R_t}{c}+T_p \right)}}\sin\mathrm{c}\left[ \left( f-\left( i-1 \right) \varDelta f' \right) T_p \right] 
\label{eq.B5}
\end{align}
where $\circledast$ denotes the convolution operation. More detailed information about convolution and Fourier transform can been seen in [\ref{cite25}]. Thereby, $\varOmega _{s}^{\left( \mathrm{SF} \right)}\left( \varDelta f,\varDelta f' \right)$ can be rewritten in (\ref{eq.B6}). 
\begin{align}
\varOmega _{s}^{\left( \mathrm{SF} \right)}\left( \varDelta f,\varDelta f' \right) =&\left. \mathcal{F} \left\{ H\left( t \right) \right\} \right|_{f=-s\varDelta f-\frac{P-1}{2}\varDelta f'}
\label{eq.B6}
\end{align}

According to the properties of sinc function, the first null of $\sin\mathrm{c}\left[ \left( f-\left( i-1 \right) \varDelta f' \right) T_p \right]$ is 
\setcounter{TempEqCnt}{\value{equation}} 
\setcounter{equation}{59}
\begin{equation}
 f_{1-\mathrm{z}}=\pm \left[ \frac{1}{T_p}+\left( i-1 \right) \varDelta f' \right] 
\label{eq.B7}
\end{equation} 
When $s>0$ and $\varDelta f\geqslant \frac{1}{Tp}\gg \varDelta f'$, 
\begin{equation}
\left| -s\varDelta f-\frac{P-1}{2}\varDelta f' \right|\gg \frac{1}{T_p}+\left( i-1 \right) \varDelta f'
\label{eq.B8}
\end{equation}
Thus, the non-diagonal elements of $\boldsymbol{D}\left( \varDelta f, \varDelta f' \right)$ in (\ref{eq.39a}) and (\ref{eq.39b}) correspond to the sidelobe energy of the sinc function, which is much smaller than the mainlobe energy corresponding to the diagonal elements.
\begin{subequations}
\begin{align}
\left.\varOmega  _{s}^{\left( \mathrm{SF} \right)}\left( \varDelta f,\varDelta f' \right) \right|_{s\ne 0}&\approx 0
\\
\left.\varOmega  _{0}^{\left( \mathrm{SF} \right)}\left( \varDelta f,\varDelta f' \right) \right|_{s= 0}&\approx PT_p
\end{align}
\label{eq.B9a}
\label{eq.B9b}
\end{subequations}

\bibsection*{REFERENCES}{}
\def\refname{}


\begin{thebibliography}{[34]}
\vspace{-0.20in}
\bibitem{bib1} J. Xu, G. Liao and H. C. So,
\newblock  Space-time adaptive processing with vertical frequency diverse array for range-ambiguous clutter suppression,
\newblock \emph{IEEE Transactions on Geoscience and Remote Sensing}, vol. 54, no. 9, pp. 5352-5364, Aug. 2016.
\label{cite1}

\bibitem{bib2} B. Huang, J. Jian, A. Basit, R. Gui and W.-Q. Wang,
\newblock  Adaptive distributed target detection for FDA-MIMO radar in Gaussian clutter without training data,
\newblock \emph{IEEE Transactions on Aerospace and Electronic Systems}, vol. 58, no. 4, pp. 2961-2972, Aug. 2022.
\label{cite2}

\bibitem{bib3}Y. Sun, W. -Q. Wang and J. Chen,
\newblock Spaceâ€“time-range clutter suppression via tensor-based STAP for airborne FDA-MIMO radar,
\newblock \emph{Signal Processing}, vol. 214, pp. 109235, Aug. 2023.
\label{cite3}

\bibitem{bib4} L. Lan, A. Marino, A. Aubry, A. De Maio, G. Liao, J. Xu and Y. Zhang,
\newblock GLRT-based adaptive target detection in FDA-MIMO radar,
\newblock \emph{IEEE Transactions on Aerospace and Electronic Systems}, vol. 57, no. 1, pp. 597-613, Feb. 2021.
\label{cite4}

\bibitem{bib5} C. -Y. Chen and P. P. Vaidyanathan,
\newblock MIMO radar space–time adaptive processing using prolate spheroidal wave functions,
\newblock \emph{IEEE Transactions on Signal Processing}, vol. 56, no. 2, pp. 623-635, Feb. 2008.
\label{cite5}

\bibitem{bib6} R. Gui, W.-Q. Wang, C. Cui and H. C. So,
\newblock Coherent pulsed-FDA Radar receiver design with time-variance consideration: SINR and CRB analysis,
\newblock \emph{IEEE Transactions on Signal Processing}, vol. 66, no. 1, pp. 200-214, Dec. 2017.
\label{cite6}

\bibitem{bib7} Z. Liu, S. Zhu, J. Xu, X. He, K. Duan and L. Lan,
\newblock  Range-ambiguous clutter suppression for STAP-based radar With Vertical Coherent Frequency Diverse Array,
\newblock \emph{IEEE Transactions on Geoscience and Remote Sensing}, vol. 61, pp. 1-17, Jul. 2023.
\label{cite7}

\bibitem{bib8} J. Xu, S. Zhu and G. Liao,
\newblock Space-time-range adaptive processing for airborne radar systems,
\newblock \emph{IEEE Sensors Journal}, vol. 15, no. 3, pp. 1602-1610, Jan. 2015.
\label{cite8}

\bibitem{bib9} S. Zhu, G. Liao, Y. Qu and Z. Zhou,
\newblock Space-time-range three dimensional adaptive processing,
\newblock In \emph{Proc. 2009 IEEE International Conference on Acoustics, Speech and Signal Processing (ICASSP)}, Taipei, China, Apr. 2009.
\label{cite9}

\bibitem{bib10} C. Wen, Y. Huang, J. Peng, J. Wu, G. Zheng and Y. Zhang,
\newblock Slow-time FDA-MIMO technique with application to STAP radar,
\newblock \emph{IEEE Transactions on Aerospace and Electronic Systems}, vol. 58, no. 1, pp. 74-95, Jun. 2021.
\label{cite10}

\bibitem{bib11} M. K. McDonald and D. Cerutti-Maori,
\newblock  Coherent radar processing in sea clutter environments, part 1: modelling and partially adaptive STAP performance,
\newblock \emph{IEEE Transactions on Aerospace and Electronic Systems}, vol. 52, no. 4, pp. 1797-1817, Aug. 2016.
\label{cite11}

\bibitem{bib12} M. K. McDonald and D. Cerutti-Maori,
\newblock  Coherent radar processing in sea clutter environments, part 2: adaptive normalised matched filter versus adaptive matched filter performance,
\newblock \emph{IEEE Transactions on Aerospace and Electronic Systems}, vol. 52, no. 4, pp. 1818-1833, Aug. 2016.
\label{cite12}

\bibitem{bib13} H. D. Joshi, R. Kaur, A. K. Singh, A. Mishra,
\newblock  An improved method for deceptive jamming against synthetic aperture radar,
\newblock \emph{International Journal of Microwave and Wireless Technologies}, vol. 10, no. 1, pp. 115-121, Feb. 2018.
\label{cite13}

\bibitem{bib14} D. Li, J. Liu, J. Li, L. Gao, P. Zheng and Z. He,
\newblock  Simulation and analysis of repeater deceptive jamming to SAR based on shift-frequency,
\newblock In \emph{Proc. 2022 International Applied Computational Electromagnetics Society Symposium (ACES-China)}, Xuzhou, China, 2022.
\label{cite14}

\bibitem{bib15} H. Bang, W.-Q. Wang, S. Zhang and Y. Liao,
\newblock FDA-based space-time-frequency deceptive jamming against SAR imaging,
\newblock \emph{IEEE Transactions on Aerospace and Electronic Systems}, vol. 58, no. 3, pp. 2127-2140, Jun. 2022.
\label{cite15}

\bibitem{bib16} W. J. Kerins,
\newblock Analysis of towed decoys,
\newblock \emph{IEEE Transactions on Aerospace and Electronic Systems}, vol. 29, no. 4, pp. 1222-1227, Oct. 1993.
\label{cite16}

\bibitem{bib17} S. Li, Z. Zong and Y. Feng,
\newblock A novel towed jamming suppression with FDA-MIMO radar,
\newblock \emph{2021 IEEE Radar Conference}, Atlanta, GA, USA, May. 2021, pp. 1-6
\label{cite17}

\bibitem{bib18} C. Dong and X. Chang,
\newblock  A novel scattered wave deception jamming against three channel SAR GMTI,
\newblock \emph{IEEE Access}, vol. 6, pp. 53882-53889, Sep. 2018.
\label{cite18}

\bibitem{bib19} M. Soumekh,
\newblock  SAR-ECCM using phase-perturbed LFM chirp signals and DRFM repeat jammer penalization,
\newblock \emph{IEEE Transactions on Aerospace and Electronic Systems}, vol. 42, no. 1, pp. 191-205, Jan. 2006.
\label{cite19}


\bibitem{bib20} L. Lan, M. Rosamilia, A. Aubry, A. De Maio and G. Liao,
\newblock FDA-MIMO transmitter and receiver optimization,
\newblock \emph{IEEE Transactions on Signal Processing}, early access, Feb, 2024, doi: 10.1109/TSP.2024.3366438.
\label{cite20}

\bibitem{bib21}C. Wen, J. Peng, Y. Zhou and J. Wu,
\newblock Enhanced three-dimensional joint domain localized STAP for airborne FDA-MIMO radar under dense false-target jamming scenario,
\newblock \emph{EEE Sensors Journal}, vol. 18, no. 10, pp. 4154-4166, May 2018.
\label{cite21}

\bibitem{bib22} L. Lan, J. Xu, G. Liao, Y. Zhang, F. Fioranelli and H. C. So,
\newblock Suppression of mainbeam deceptive jammer with FDA-MIMO radar,
\newblock \emph{IEEE Transactions on Vehicular Technology}, vol. 69, no. 10, pp. 11584-11598, Oct. 2020.
\label{cite22}

\bibitem{bib23} L. Lan, M. Rosamilia, A. Aubry, A. De Maio, G. Liao and J. Xuï¼Œ
\newblock Adaptive target detection with polarimetric FDA-MIMO radar,
\newblock \emph{IEEE Transactions on Aerospace and Electronic Systems}, vol. 59, no. 3, pp. 2204-2220, Sep. 2022.
\label{cite23}

\bibitem{bib24}G. Wang and Y. Lu,
\newblock Clutter rank of STAP in MIMO radar with waveform diversity,
\newblock \emph{IEEE Transactions on Signal Processing}, vol. 58, no. 2, pp. 938-943, Feb. 2010.
\label{cite24}

\bibitem{bib25} B. B. Mahafza,
\newblock Radar system analysis and design using MATLAB (4th Edition),
\newblock Taylor \& Francis Group, New York, pp. 690, Mar. 2022.
\label{cite25}

\bibitem{bib26}K. Zhou, D. Li, Y. Su and T. Liu,
\newblock Joint design of transmit waveform and mismatch filter in the presence of interrupted sampling repeater jamming,
\newblock \emph{IEEE Signal Processing Letters}, vol. 27, pp. 1610-1614, Sep. 2020.
\label{cite26}

\bibitem{bib27}W. Wu, J. Zou, J. Chen, S. Xu and Z. Chen,
\newblock False-target recognition against interrupted-sampling repeater jamming based on integration decomposition,
\newblock \emph{IEEE Transactions on Aerospace and Electronic Systems}, vol. 57, no. 5, pp. 2979-2991, Oct. 2021.
\label{cite27}

\bibitem{bib28}W. Mao, H. Wang, S. Zhang and X. Liu,
\newblock  A novel deceptive jamming method via frequency diverse array,
\newblock In \emph{Proc. IGARSS 2019 - 2019 IEEE International Geoscience and Remote Sensing}, Symposium, Yokohama, Japan, Nov. 2019, pp. 2369-2372.
\label{cite28}

\bibitem{bib29}S. Li, Z. Zong and Y. Feng,
\newblock  A novel towed jamming suppression with FDA-MIMO radar,,
\newblock In \emph{Proc. 2021 IEEE Radar Conference (RadarConf21)}, Atlanta, GA, USA, Jun. 2021, pp. 1-6.
\label{cite29}

\bibitem{bib30}L. Wang, W. -Q. Wang and H. C. So,
\newblock  Covariance matrix estimation for FDA-MIMO adaptive transmit power allocation,
\newblock \emph{IEEE Transactions on Signal Processing}, vol. 70, pp. 3386-3399, Jun. 2022.
\label{cite30}

\bibitem{bib31}Z. Pan, Y. Li, S. Wang and Y. Li,
\newblock  Joint optimization of jamming type selection and power control for countering multifunction radar based on deep reinforcement learning,
\newblock \emph{IEEE Transactions on Aerospace and Electronic System}, vol. 59, no. 4, pp. 4651-4665, Aug. 2023.
\label{cite31}

\bibitem{bib32}J. Xu, G. Liao, Y. Zhang, H. Ji and L. Huang,
\newblock  An adaptive range-angle-Doppler processing approach for FDA-MIMO radar using three-dimensional localization,
\newblock \emph{IEEE Journal of Selected Topics in Signal Processing}, vol. 11, no. 2, pp. 309-320, March 2017.
\label{cite32}

\bibitem{bib33}Q. Qu, Y. -L. Wang, W. Liu, S. Wei and Q. Du,
\newblock  IRNet: Interference recognition networks for automotive radars via autocorrelation features,
\newblock \emph{IEEE Transactions on Microwave Theory and Techniques}, vol. 70, no. 5, pp. 2762-2774, May 2022.
\label{cite33}

\bibitem{bib34} A. Hassanien and S. A. Vorobyov,
\newblock Phased-MIMO radar: a tradeoff between phased-array and MIMO radars,
\newblock \emph{IEEE Transactions on Signal Processing}, vol. 58, no. 6, pp. 3137-3151, Feb. 2010.
\label{cite34}

\end{thebibliography}
\end{document}